**Metal insulator transitions in perovskite SrIrO$_3$ thin films**

Abhijit Biswas[1], Ki-Seok Kim[1,2], and Yoon Hee Jeong[1,a)]

[1]*Department of Physics, POSTECH, Pohang, 790-784, South Korea*
[2]*Institute of Edge of Theoretical Science (IES), POSTECH, Pohang, 790-784, South Korea*

*Abstract*

Understanding of metal insulator transitions in a strongly correlated system, driven by Anderson localization (disorder) and/or Mott localization (correlation), is a long standing problem in condensed matter physics. The prevailing fundamental question would be how these two mechanisms contrive to accomplish emergent anomalous behaviors. Here we have successfully grown high quality perovskite SrIrO$_3$ thin films, containing a strong spin orbit coupled 5*d* element Ir, on various substrates such as GdScO$_3$ (110), DyScO$_3$ (110), SrTiO$_3$ (001) and NdGaO$_3$ (110) with increasing lattice mismatch, in order to carry out a systematic study on the transport properties. We found that metal insulator transitions can be induced in this system; by either reducing thickness (on best lattice matched substrate) or changing degree of lattice strain (by lattice mismatch between film and substrates) of films. Surprisingly these two pathways seek two distinct types of metal insulator transitions; the former falls into disorder driven *Anderson* type whereas the latter turns out to be of unconventional *Mott-Anderson* type with the interplay of disorder and correlation. More interestingly, in the metallic phases of SrIrO$_3$, unusual non-Fermi liquid characteristics emerge in resistivity as $\varDelta\rho \propto T^\varepsilon$ with $\varepsilon$ evolving from 4/5 to 1 to 3/2 with increasing lattice strain. We discuss theoretical implications of these novel phenomena to shed light on the metal insulator transitions.

[a] Author to whom correspondence should be addressed: yhj@postech.ac.kr

# I. INTRODUCTION

Ruddlesden-Popper phases of strontium iridates $Sr_{n+1}Ir_nO_{3n+1}$ ($n$ = 1, 2, and ∞), have been a subject of active investigations in recent years as the interplay between local Coulomb interaction and strong spin orbit coupling (SOC) gives rise to rich phase diagrams including metal insulator transitions (MITs) [1,2]. Since the SOC of an element is proportional to $z^4$ with $z$ being the atomic number, the strength of SOC in 5$d$ Ir oxide is as large as ~0.5 eV; this large SOC then would effectively reduce the conduction bandwidth ($W$) and make it comparable to the Coulomb interaction ($U$) [3]. Layered $Sr_2IrO_4$ compound ($n$ = 1) is a good example among the Ruddlesden-Popper series $Sr_{n+1}Ir_nO_{3n+1}$, which was demonstrated to be a $J_{eff}$ = 1/2 antiferromagnetic Mott insulator [4]. $Sr_3Ir_2O_7$ ($n$ = 2), on the other hand turns out to be barely an insulator which displays a ferromagnetic state with a Curie temperature of 285 K [5]. With increasing the number of $IrO_2$ planes, that is with increasing $n$, the bandwidth of Ir 5$d$ states becomes broader and as a result, perovskite $SrIrO_3$ ($n$ = ∞) becomes a correlated metal [1]. As a matter of fact, perovskite $SrIrO_3$ would be a correlated bad metal, i.e. mean free path comparable to the inter-atomic distance, and is presumably close to a metal-insulator phase boundary. It is thus natural to expect that perovskite $SrIrO_3$ would be susceptible to external perturbations and perhaps an appropriate perturbation on this system would be able to induce a MIT, which is one of the fundamental problems in condensed matter physics but remains not fully understood till date [6-8].

In real physical systems, disorder is another important factor determining transport properties and it is generally stated that the two most important mechanisms underlying MITs in correlated systems are interaction or correlation ($U$) driven Mott localization and disorder ($D$) driven Anderson localization. The MIT problem becomes particularly difficult in the simultaneous presence of both *U* and *D*. While previous efforts have clarified the issue in the weak limits of *U* or *D* becoming small; it has been a long standing puzzle how correlation (disorder) becomes renormalized in the presence of strong disorder (correlation) [9,10]. From the experimental point of view, various aspects of MITs are frequently studied in bulk samples; single crystalline as well as polycrystalline one, by modifying them with chemical pressure, change in stoichiometry, etc. With thin films where metastable phases can often be stabilized, however, one would take advantages over bulk ones because key parameters such as bandwidth (*W*), correlation *(U)*, disorder *(D)*, and the strength of SOC can be changed to some extent by varying film thicknesses or lattice strains utilizing different lattice mismatched substrates. Then one may hope that the study of MIT in perovskite $SrIrO_3$ thin films would also be able to provide new insights into the problem.

In view of these facts, we resorted to thin film growth techniques to obtain metallic $SrIrO_3$ in perovskite form and also to induce new physical phenomena (MIT and others if any) different from bulk properties. In this article we wish to report the results of our systematic investigations on the transport properties of perovskite $SrIrO_3$: (1) reducing thickness (dimensionality) of the films from 35 nm to 3 nm leads to a MIT of *Anderson* type, where interactions do not seem to play a significant role and (2) increasing the compressive strain in the films with fixed thickness of 35 nm via lattice mismatched substrates also causes a MIT, but seemingly of different type of *Mott-Anderson* one, where both correlation and disorder play a significant role. In particular, in the metallic phases of 35 nm $SrIrO_3$ films grown on different substrates, non-Fermi liquid behaviors with resistivity $\Delta\rho \propto T^\varepsilon$ with $\varepsilon < 2$ are found over a wide temperature range. Furthermore unexpected evolution of the temperature exponent $\varepsilon$ is observed before an insulating phase sets in, that is, ε changes from 4/5 to 1 to 3/2 as the strain in the films increases.

## II. EXPERIMENTAL DETAILS

While the equilibrium structure of SrIrO$_3$ at room temperature and atmospheric pressure takes that of hexagonal BaTiO$_3$, perovskite SrIrO$_3$ is known to be obtainable by applying extreme conditions of elevated pressure (40 kbar) and temperature (1000 $^o$C) and subsequent quenching [11]. Instead of applying these extreme conditions, however, we attempted to stabilize perovskite SrIrO$_3$ via different crystal growth route, namely thin film synthesis. We have successfully synthesized SrIrO$_3$ thin films of different thicknesses on four substrates such as GdScO$_3$ (110), DyScO$_3$ (110), SrTiO$_3$ (001), and NdGaO$_3$ (110) [12]. SrIrO$_3$ thin films were fabricated by pulsed laser deposition (KrF laser with λ = 248 nm) from a polycrystalline target. The laser was operated at frequency 4 Hz, and the substrate temperature and oxygen partial pressure were 550 °C and 20 mTorr, respectively. All the films were post annealed at the same growth temperature and oxygen partial pressure for 30 min to compensate for any oxygen deficiency. It may be stressed that all the substrates used in the present experiment were treated prior to growth to ensure atomically flat surfaces which are essential for high quality film growth [13-15]. It is noted that while perovskite SrIrO$_3$ is of orthorhombic structure, GdScO$_3$, DyScO$_3$, and NdGaO$_3$ are also orthorhombic but SrTiO$_3$ is cubic. Orthorhombic indices are typically used to indicate the orientation of orthorhombic substrates as in GdScO$_3$ (110), DyScO$_3$ (110), and NdGaO$_3$ (110). It is, however, more convenient for mutual comparison to use *pseudocubic* lattice parameters and cubic indices with subscript *c* attached for the orthorhombic perovskites. (We omit the orientation indices (110) or (001) for substrates hereafter.) The pseudocubic ($a_c$) lattice parameter converted from the orthorhombic lattice parameters of bulk SrIrO$_3$ is found to be ∼3.96 Å [11]. If one takes this value literally, it would match GdScO$_3$ substrates ($a_c$ = 3.96 Å) and correspond to +0.50%, +1.54%, and +2.59% compressive strain for the films on DyScO$_3$ ($a_c$ = 3.94 Å), SrTiO$_3$ ($a$ = 3.90 Å), and NdGaO$_3$ substrates ($a_c$ = 3.86 Å), respectively. The in-plane lattice parameters of the substrates are in relationship to that of perovskite SrIrO$_3$ as illustrated in Fig. **1**. It is seen that the substrates ranges conveniently from very well matched (GdScO$_3$) to strongly mismatched (NdGaO$_3$).

Extensive structural analyses of the SrIrO$_3$ films on the four substrates were carried out and the results are summarized in Fig. **2**. First of all, X-ray diffraction (XRD) measurements for the SrIrO$_3$ films of thickness 35 nm shows only the crystalline peaks without any impurity or additional peaks. For clarity, only the low angle (001)$_C$ peak has been shown in Fig. **2(a)-(d)** where thickness fringes are clearly visible. It is of interest to note that for higher strain case peak width becomes broader despite the same thickness of the films (The thicknesses of the films were confirmed by X-ray reflectivity as well as by depth profile measurements). In general the fringe oscillations around Bragg peak is a result of coherent scattering from a finite number of lattice planes with coherent thickness of the film, confirming the homogeneous nature of the surface. But while increasing the lattice strain, as x-ray profile suggests possibly fringe peaks merge with the Bragg peaks; indicating that highly strained films may have sublayers with coherent and incoherent lattice planes (indication of possible disorder effect). In order to further characterize the crystalline quality of the films, XRD ω scans (rocking curve) were performed. In Fig. **2(e)** shown are the results of the ω scan around the (001)$_C$ peak of SrIrO$_3$ films on GdScO$_3$ for varying thicknesses. In comparing the full width at half maximum (FWHM) of the rocking curves, the center position of the rocking curves were coincided. It is seen from the figure that the FWHM of the films with thickness less than 35 nm is small (< 0.05°), but that of the thick film (70 nm) is extremely broad reaching 1.357°. It was generally found that the films on all four substrates with thickness less than 40 nm maintain the sharpness in the peak of the rocking scans, and thus the maximum thickness of the films were fixed at 35 nm. A real space microstructural image of a typical 35 nm least strained $SrIrO_3$ film on best lattice matched $GdScO_3$ substrate, obtained by cross sectional transmission electron microscopy (TEM), is shown in Fig. **2(f).** It is seen that at the interface, the film adopts the same crystallographic orientation as that of the substrate revealing a high crystalline quality with well-defined atomic order.

We have also measured the strain in the films grown on four different substrates utilizing reciprocal space mapping (RSM); RSM is a two-dimensional mapping of film and substrate Bragg peaks in in-plane and out-of-plane directions, and allows the detailed structural analysis of strained films. In Fig. **2(g)** shown are the RSM results around the $(103)_C$ reflection peak of the $SrIrO_3$ films of thickness 35 nm on $GdScO_3$, $DyScO_3$, $SrTiO_3$, and $NdGaO_3$ substrates. It is seen that while in-plane $Q_x$ of the films and that of the substrates are locked together, out-of-plane $Q_z$ of the films decreases as the lattice mismatch increases. This is of course due to the fact that as the in-plane compressive strain increases with lattice mismatch, the out-of-plane lattice expansion occurs correspondingly. From the RSM data, the out-of-plane lattice parameters of the $SrIrO_3$ films were determined to be 3.99 Å, 4.01 Å, 4.03 Å, and 4.06 Å for the $GdScO_3$, $DyScO_3$, $SrTiO_3$, and $NdGaO_3$ substrates, respectively. Table **I** is the summary of the structural analyses of the $SrIrO_3$ films on the four substrates along with the approximate bulk lattice parameters. It is seen from the table that the films grown on $GdScO_3$ represent the most natural state of perovskite $SrIrO_3$ available and that in-plane compression brings about overall compression despite of out-of-plane expansion. It may also be pointed out that while the out-of-plane lattice parameter of the film grown on GdScO$_3$ appears to increase from the bulk value, this probably means that the pseudocubic lattice parameters of bulk SrIrO$_3$ are approximate ones.

In order to check the homogeneity in the distribution of cationic elements within the films, chemical analysis employing secondary ion mass spectrometry (SIMS) was attempted. The SIMS results of the SrIrO$_3$ films deposited on GdScO$_3$, DyScO$_3$, and SrTiO$_3$ are shown in Fig. **2(h)-(j)**. Generally the elemental distributions on these substrates appear to be homogeneous at the resolution level of SIMS measurements. For the films grown on NdGaO$_3$, however, we observe an inhomogeneous distribution of Ir along the thickness direction as shown in Fig. **2(k)**. At the substrate-film interface Ir deficiency was found with Ir surplus at the surface. This inhomogeneity, evident only for the case with the largest lattice mismatch among the films, seems to be a release mechanism of accumulating strain energy in the film consistent with the fact in XRD the film was not found to be fully homogeneous; indicating the presence of possible disorder. The morphology of the films were also measured by atomic force microscopy (AFM) as shown in the insets of Fig. **2(h)-(k)**. The AFM images confirm that the surfaces are flat within roughness of ~1 nm. Measurements of electrical transport properties of SrIrO$_3$, the main theme of the present work, were performed using the four-probe van der Pauw geometry, where Au electrodes were deposited at the four corners of the samples to make contacts.

### III. METAL INSULATOR TRANSITIONS IN SrIrO$_3$

**III. 1. The most natural state: perovskite SrIrO$_3$ on GdScO$_3$**

In studying the transport properties of SrIrO$_3$, it would be most reasonable to start with bulk single crystals. Perovskite SrIrO$_3$ single crystals however, are not available yet, so that we start with 35 nm thick SrIrO$_3$ films deposited on the best lattice matched GdScO$_3$ substrates. It was already noted that this would be the most natural state of perovskite SrIrO$_3$ available. Figure **3(a)** shows the resistivity behaviors of a 35 nm film on GdScO$_3$ substrate. First of all, it is noted that the resistivity at room temperature is ρ = 1.4 mΩ·cm, which is close to but smaller than the bulk polycrystals value, ∼2.5 mΩ·cm at 300 K [16]. The 35 nm thin film was found to remain metallic (*dρ/dT* > 0) down to the lowest temperature (300 mK) measured; no kink or peak in resistivity indicating a phase transition was observed in the whole temperature range. The most surprising feature about the transport behaviors of the SrIrO$_3$ film on GdScO$_3$ is that its electrical resistivity exhibits a $T^{4/5}$ dependence in the wide range of temperature, from ~10 K to 300 K, as evidenced in Fig. **3(b)**. Fitting the experimental data to $\rho(T) = \rho_0 + A \cdot T^{4/5}$ over the temperature range from 10 K to 300 K, we obtain the residual resistivity $\rho_0 = 0.969$ and $A = 0.00436$ in proper units with resistivity in mΩ·cm and temperature in K. We may emphasize that the residual resistivity ~1 mΩ·cm is smaller than so-called Mott-Ioffe-Regel (MIR) limit (see the next section) [17]. The *sublinearity-in-temperature* in electrical resistivity over such an extended temperature range in a correlated metallic system is quite remarkable and should be regarded as one of the major findings in perovskite SrIrO$_3$ films. (We will come back to this point in the conclusion section.) In a phonon dominated region at high temperatures, for example, a linear temperature dependence of resistivity would be observed in metals unless there are other reasons such as spin fluctuation. Below 10 K, the Fermi liquid behavior of electrical resistivity, $\rho \propto T^2$, is found to hold down to 300 mK as shown in inset of Fig. **3(b)**. This kind of crossover from Fermi liquid behaviors at low temperatures to non-Fermi liquid ones at higher temperatures is typically seen in strongly correlated systems [18].

To get further insight on the electronic nature of $SrIrO_3$, we performed Hall Effect measurements and estimated the carrier concentration ($n$) and mobility ($\mu$). The results are shown in Fig. **4(a)**; the obtained value of the carrier concentration is $n \approx 10^{20}$ cm$^{-3}$ while the Hall mobility is low over the whole range of temperature with $\mu \approx 10$ cm$^2$V$^{-1}$s$^{-1}$ exhibiting semimetallic-like nature. Additionally, field dependence resistance (Fig. **4(b)**) shows no significant change in the slope except increase of resistance at low T indicating no significant role of spin fluctuations. Figure **4(c)** displays the magnetoresistance (MR) at 5 K, defined as $\frac{\Delta\rho}{\rho_0} = \frac{\rho(B)-\rho(0)}{\rho(0)}$, with applied field up to $\pm 9$ Tesla. Note that both in-plane and out-of-plane MRs were measured. In the out-of-plane configuration, the current and the field were naturally perpendicular; in the in-plane configuration, where both the current and the field are in the film plane, the applied magnetic field was still perpendicular to the current direction. Thus, both MRs are transverse. It is seen from Fig. **4(c)** that both in-plane and out-of-plane MRs are positive and proportional to $B^2$. This almost isotropic MR indicates that the mean free path is sufficiently smaller than the thickness, and thus the film is of three dimensional natures at 35 nm. The positive quadratic MR particularly in the Fermi liquid regime, is similar to the one found in normal metallic systems, and may be attributed to the Lorentz contribution [19]. Based on these experimental measurements, we may conclude that 35 nm thin films of $SrIrO_3$ deposited on $GdScO_3$ are a three dimensional correlated metal.

**III. 2. Metal-insulator transition driven by disorder: thickness variation of $SrIrO_3$ films**

**on GdScO$_3$**

Having secured a three dimensional metallic phase of SrIrO$_3$, we wish to explore MIT occurring in this correlated system. Reducing thickness in many metallic correlated systems brings about interesting quantum behaviors such as theoretically predicted high T$_C$ superconductivity, topological insulators, and apparent metal insulator transitions [20-22]. In order to check whether a MIT occurs in SrIrO$_3$ as a result of thickness reduction, we systematically reduce the thickness of the films on GdScO$_3$ substrates. Figure **5(a)** shows the measured sheet resistance ($R_S$) of the SrIrO$_3$ films, as a function of temperature, for various thicknesses. As the thickness is reduced from 35 nm to 10 nm and further to 4 nm, the films still remains metallic. For the 4 nm film, however, a strong upturn appears in the sheet resistance below 15 K as indicated by an arrow in the figure. (Also see Fig. **6(c)**.) This kind of low temperature upturns in resistivity is often seen when a system approaches MIT from the metallic side. On further reducing the thickness to 3 nm, the film indeed becomes fully insulating and thus a MIT is identified at thickness between 4 nm and 3 nm. Although the thickness dependent MIT was previously observed in several thin film materials but the fundamental thickness limit of the MIT has been an issue of some debate and seems to depend on the quality of samples [23]. In the case of SrIrO$_3$, the observed 3 nm (～7.5 unit cells) thickness limit for the MIT appears to be in a similar range to those found in other systems [24].

To specify the origin of the MIT occurring with reducing the thickness in SrIrO$_3$, attention is drawn to the fact that the low temperature resistance upturn in 4 nm thin film is explained well by two dimensional *weak localization,* due to enhanced quantum backscattering interference of electronic waves diffusing around scattering centers, where conductivity would show the characteristic ln *T* behavior [6]. Figure **5(b)** illustrates that the sheet conductance (inverse of sheet resistance) in the low temperature upturn region indeed exhibits the ln *T* variation. Moreover, for the fully insulating phase of 3 nm film, the measured data can be fitted at low temperatures with the variable range hopping (VRH) model due to Mott, i.e., ln $\sigma \propto 1/T^{1/(d+1)}$ where $\sigma$ is the conductivity and *d* the dimension. [7] Fig. **5(c)** shows that VRH with *d* = 2 holds for the 3 nm film up to approximately 15 K. Thus, it appears to be Anderson localization due to disorder which brings about a MIT with reduction of thickness; in good agreement with other studies [12]. It certainly makes sense and is well known that the disorder in the $SrIrO_3$ film, of whatever origin, becomes more effective in scattering charge carriers while lowering thickness. In this view of MIT, the transition should occur when electron mean free path *l* becomes less than or equal to lattice spacing ($l \leq a$) and simple kinetic theory of quasiparticles breaks down. In such a situation product of $k_Fl$ becomes close to unity where $k_F$ is the Fermi wave number, and the sheet resistance should cross the so-called Mott-Ioffe-Regel (MIR) limit of quantum resistance, $h/e^2 \approx 26$ k$\Omega$ [17]. It is easily seen that the sheet resistance value for the $SrIrO_3$ films reaches the MIR limit at 3 nm as shown in Fig. **5(a)**.

Although Anderson localization due to disorder explains coherently the thickness dependent MIT for $SrIrO_3$, there may be another possibility, a bandwidth controlled MIT occurring with reducing thickness [25]. In this scenario of bandwidth controlled MIT, decreasing thickness would cause reduction in atomic coordination, which in turn decreases bandwidth *W* and increases effective correlation (*U/W*). Enhanced effective correlation then opens up a correlation gap and makes the system insulating [7]. For this scenario to hold, it may be pointed out, the continuous change in the effective interaction, which would eventually cause a MIT, should bring about a corresponding change in the metallic regime as well. Figure **6** tests this point; the sheet resistance of the films in the metallic region is shown in Fig. **6(a)-(c)**. It is seen from the figure that with changing thickness of the films, the $T^{4/5}$ dependence of the sheet resistance remains the same; thus, it appears that correlation effects are not greatly enhanced with reduced thickness. In contrast to this situation, we shall encounter a genuine bandwidth varying case with the corresponding exponent $\varepsilon$ change in the next section when we deal with films with varying strains.

To further ascertain the localization origin, the MR, $\frac{\Delta R}{R_0} = \frac{R(B)-R(0)}{R(0)}$, for the 4 nm sample at various temperatures is shown in Fig. **6(d)**. It is striking that while the MR at 20 K, where the $T^{4/5}$ dependence still holds, is positive, it becomes negative at 5 K where the resistivity upturn persists. This negative MR should be contrasted to the positive MR found at the same temperature in the 35 nm film as shown earlier in Fig. **4(c)**. Remembering that the system is in the weak localization regime as illustrated in Fig. **5(b)**, we conclude that negative MR is caused by the magnetic field suppressing the localization interference of back scattered waves. In the metallic side (20 K) above the weak localization regime, on the other hand, MR is positive. At an intermediate temperature 10 K, it is seen that MR is almost cancelled by the two opposing tendencies. Thus, all the electrical transport measurements for the SrIrO$_3$ films of different thicknesses on GdScO$_3$ substrates converge on the conclusion that the thickness dependent MIT in this system is of *Anderson* type driven by disorder.

**III. 3. Metal insulator transition driven by bandwidth reduction: SrIrO$_3$ films on different substrates**

Now that we have established the disorder driven MIT in SrIrO$_3$ films by thickness variation, let's move on to another direction in exploring MITs of SrIrO$_3$. This time we shall fix the thickness of films and change the underlying substrates and thereby control the lattice strain. Note that this approach represents one of the advantages for dealing with thin film samples and strain engineering is a classic route for controlling functional properties of thin films [26]. To realize this idea, SrIrO$_3$ thin films of thickness 35 nm were deposited on GdScO$_3$, DyScO$_3$, SrTiO$_3$, and NdGaO$_3$ substrates under the same deposition conditions; the thickness 35 nm was chosen to ensure that the system remains three dimensional. Generally for ABO$_3$ perovskites, an externally imposed strain would cause a change of the B-O-B bond angle as well as bond length, which in turn would induce a bandwidth change as a major effect as $W \propto \frac{\cos \varphi}{d^{3.5}}$, where *d* is the B-O bond length and $\varphi = (\pi - \theta)/2$ is the buckling deviation of the B-O-B bond angle *θ* from π [27]. For perovskite thin films it was reported that in-plane strain is mostly adopted via the tilting of BO$_6$ octahedra or a change in *θ* rather than a change in *d* [28]. Considering the biaxial nature of the strain in the film, however, the relationship between the strain in the film and its bandwidth may not be so straightforward. Nevertheless, the average values for *θ* and *d* would be sufficient in getting the general trend. Then the overall compression existing in the films, given in Table **I**, would result in corresponding bandwidth reduction. It is pointed out again that MIT in a correlated system is controlled by competition among relevant energy scales such as *W, U, D,* etc.; and important parameters governing the physics are the effective correlation (*U/W*), effective disorder (*D/W*), etc.; thus, bandwidth reduction would have far-reaching effects in a correlated system [9]. Perovskite SrIrO$_3$ is a correlated system, in particular, close to the metal insulator phase boundary, and it is natural to expect that changing strain in $SrIrO_3$ would bring about corresponding interesting changes in transport properties.

Figure **7** shows the measured electrical resistivity of $SrIrO_3$ films, as a function of temperature, deposited on the four substrates. The most conspicuous feature in the figure is obviously the appearance of a new MIT in the $SrIrO_3$ system; while the thin films on $GdScO_3$, $DyScO_3$, and $SrTiO_3$ show metallic behaviors, the one on $NdGaO_3$ is insulating. Note that the insulating phase appears in the film with the largest strain is due to its largest lattice mismatch with the substrate. For the films on $DyScO_3$ and $SrTiO_3$ with intermediate strains, they still maintain metallicity but with the resistivity upturns at 20 K and 50 K (arrows in the figure) respectively, which continue to increase with decreasing temperature. The MIT in $SrIrO_3$ films under strain is obviously related to the change in bandwidth (*W*) arising from $IrO_6$ octahedral tilting as explained above; however, the identification of the mechanism of how this bandwidth change brings about the MIT would require more detailed quantitative analysis of the data. It is pointed out here that a MIT has also been observed previously in $Ca_{1-x}Sr_xIrO_3$ thin films on $GdScO_3$ (110) substrates via A-site doping or controlling chemical pressure [29]. It is of interest to note that the MIT in these compounds with strong SOC has been interpreted to be caused by a change in band width (*W*) arising from a variation in octahedral distortion with doping. Detailed quantitative analysis of the transport behaviors of this system, however, was not attempted.

In order to gain a clue about the mechanism of the strain dependent MIT, we focus on the nature of transport in the metallic phases and try to fit the temperature dependence of the resistivity of the films to $\rho(T) = \rho_0 + A \cdot T^\varepsilon$, where $\varepsilon$ would identify scattering processes. In the conventional Fermi liquid framework, for example, electron-electron interaction at low temperatures gives rise to $\varepsilon = 2$, and at high temperatures above the Debye temperature dominant electron-phonon scattering gives rise to $\varepsilon = 1$. First of all, we recall the transport behaviors of the 35 nm SrIrO$_3$ film grown on GdScO$_3$: the surprising $T^{4/5}$ dependence from 10 K - 300 K and a Fermi liquid behavior ($\rho \propto T^2$) below 10 K down to 300 mK. For the ease of comparison, these behaviors are displayed again in Fig. **8(a)**. On the other hand, the films on DyScO$_3$ and SrTiO$_3$ with increasing lattice mismatch and corresponding compressive strain follows power law of $\rho \propto T$ and $\rho \propto T^{3/2}$ in the metallic regions respectively, as shown in Fig. **8(b)** and **(c)**. For these two samples, resistivity upturns appear at 20 K and 50 K, respectively, and continue to increase with decreasing temperature; these upturns appear again, as in the thickness variation case, at low temperatures upon approaching the MIT from the metallic side.

The continuous evolution of the power exponent $\varepsilon$ from 4/5 to 1 to 3/2 found in the strain varying case is in striking contrast to the previous thickness variation case where $\varepsilon$ stays constant at sublinear 4/5 throughout. It thus seems to suggest that disorder alone, which was the main physics for the thickness variation case, would not be sufficient to account for the strain varying situation and an additional piece (or pieces) of physics be needed. To this end, it is recalled again that reduction in bandwidth $W$, which arises from compressive strain, brings forth enhancements in relevant parameters: effective disorder ($D/W$) as well as effective correlation ($U/W$). Thus, we may anticipate a very likely possibility that interplay of enhanced correlation and disorder due to the bandwidth reduction would lead to the variation of the temperature exponent in resistivity. Keeping this in mind for now, however, we may check another possibility, namely the spin fluctuations. In metals near an antiferromagnetic quantum critical point, for example, it was shown theoretically that interplay of anisotropic scattering due to spin fluctuations and isotropic impurity scattering could lead to a variation in power law exponent of resistivity depending on the amount of disorder [30]. It was already stated in the previous section that we did not observe any unusual features related with spin fluctuations in the in-plane and out-of-plane MR measurements for the SrIrO$_3$ films grown of GdScO$_3$ (see Fig. 4**(b)** and **(c)**). We carried out magnetic measurements for the films on all substrates too but did not found any evidence indicating a magnetic signature. Thus, we conclude that the system still remains truly a paramagnetic correlated metal without strong spin fluctuations, and let's turn our attention to the insulating side.

For the insulating one on NdGaO$_3$, the resistivity below approximately 35 K is fitted well with the three dimensional VRH model (ln $\sigma \propto 1/T^{1/4}$) as shown in Fig. **8(d)**. The validity of the VRH model indicates that disorder plays a significant role also in this film; the SIMS data of Fig. **2(k)** already revealed the presence of cationic disorder in the system. For further insight of MIT, the low temperature upturn regions of the resistivity of the metallic films on DyScO$_3$ and SrTiO$_3$ substrates are analyzed; the insets of Fig. **8(b)** and **(c)** show that the sheet conductance (inverse of resistance) is proportional to the logarithm of *T*. Since the logarithmic temperature dependence is usually regarded as the hall mark of two dimensional weak localization as opposed to the three dimensional $\sqrt{T}$ dependence, [6] we seem to have a discrepancy, that is, it is difficult to understand an appearance of the logarithmic behavior in the three dimensional metallic films. In addition, this logarithmic behavior seems to evolve into the physics of three dimensional VRH in the case of the film on NdGaO$_3$. Now if these upturns in resistivity are really due to weak localization, then we would expect similar negative MR as already seen in Fig. **6(d)** of the 4 nm film on GdScO$_3$.

To resolve this issue, we again measured out of plane MR of all four films on GdScO$_3$, DyScO$_3$, SrTiO$_3$, and NdGaO$_3$ substrates. Figure **9(a)-(d)** display the results, MR with a field up to ±9 Tesla. Surprisingly, we observe that the MR of all the films is always positive and proportional to $B^2$. First of all, it is clear that the positive MR of the two cases, the metallic one on GdScO$_3$ and the insulating one on NdGaO$_3$, would have different origins even though they look similar. The positive MR in the former case was attributed to the Lorentz contribution; the positive MR in the latter case is the effect occurring in the already localized state, where such effect as the shrinkage of the wave function was suggested to account for positive MR [31]. For the two films with intermediate strains, an appearance of positive MR indicates again, despite of the ln $T$ behaviors in the upturn region, that disorder alone cannot fully account for the phenomena. In contrast, the ln $T$ behavior of the two dimensional film on GdScO$_3$ was accompanied by negative MR in Fig. **6(d)**. Recognizing the seemingly contradicting features such as the evolution of temperature exponent in the metallic conductivity, the low temperature ln $T$ upturn behaviors, and positive MR, we come to conclude that not only disorder but also electron-electron correlation should be taken into account to be able to ascertain the strain dependent MIT (calling *Mott-Anderson* type) of SrIrO$_3$. Full theory incorporating these two ingredients simultaneously is difficult to establish and at the moment it is still lacking.

## IV. CONCLUSION AND DISCUSSION

In summary, we have established the following: (1) high quality 35 nm thin film of perovskite SrIrO$_3$ on GdScO$_3$ substrate is a correlated paramagnetic metal exhibiting non-Fermi liquid behavior in electrical resistivity $\Delta\rho \propto T^\varepsilon$ with exponent $\varepsilon = 4/5$. (2) Reducing thickness from 35 nm to 3 nm, SrIrO$_3$ films undergoes a MIT of *Anderson* type driven by disorder. (3) With an increase of the imposed compressive strain via lattice mismatched substrates, SrIrO$_3$ films also undergoes a MIT but seemingly of different nature, presumably *Mott-Anderson* type, where correlation does play a significant role in the presence of disorder. In particular, the temperature exponent $\varepsilon$ of electrical resistivity evolves with strain from 4/5 to 3/2 before the system goes into the insulating phase.

Although we have revealed the different kinds of MITs from transport measurements and subsequent empirical data analyses in SrIrO$_3$ films, understanding of the microscopic mechanism underlying the MITs would require further efforts on the theoretical front. Generally speaking, the qualitatively different MITs we observed in the two cases of SrIrO$_3$ thin films would result from the fact that they follow different evolution paths in the global phase diagram of, for example, a disordered Hubbard model on the plane of interaction (*U)* and disorder (*D)* scaled with bandwidth (*W)* [10]. It would be a challenging task to construct a proper theory which accounts for the experimental results from a minimal model. Again on general grounds, the observed evolution of power law exponents as well as MIT could be sorted out in the framework of dynamical mean field theory (DMFT) as the system does not show any long range magnetic ordering but may need local moments to account for the experimental results [32]. Also important for SrIrO$_3$ is the fact that it has large SOC, and this fact should be taken into account. Recent theoretical calculations showed that various ground states emerge for bulk SrIrO$_3$ as both Hubbard *U* and the strength of SOC are tuned [33]. It is perhaps desirable to pursue such theoretical calculations with varying degrees of disorder.

Mention should be made on one particular theoretical effort which attempts to vary the strength of disorder in a correlated system. A previous study shows that if strength of disorder crosses a certain limit ($D > D_{NFL}$), the system enters a "*Griffiths phase*" displaying metallic non-Fermi liquid behaviors and even with stronger disorder above some critical value $D_C$ ($D > D_C > D_{NFL}$) the system undergoes a MIT [34]. Currently, we are exploring a possibility that the present strain dependent MIT may be described as a Mott-Anderson transition occurring in the presence of strong disorder, perhaps calling MIT of *Mott-Anderson-Griffiths* type. Our approach based on this scenario is sketched in the *Supplementary Material*. [35] The idea is that this correlated disordered system may show a power law distribution for the Fermi liquid coherence temperature, which originates from inhomogeneous formation of Mott insulating islands and Fermi liquid metal islands. This power law distribution is essential for the Griffiths type scenario. The power law exponent for electrical resistivity then would depend on both interaction and disorder, and this may provide a possible explanation why strain dependent non-Fermi liquid exponents have been observed in $SrIrO_3$.

On the experimental front, while we revealed many new properties in the present work including MITs in $SrIrO_3$, but we also raised many questions which require subsequent efforts. The foremost task we face would be to reduce the amount of disorder in the films as much as possible. The residual resistivity of the most natural film (35 nm one on $GdScO_3$) is about 1 mΩ·cm; this value is smaller than the Mott-Ioffe-Regel limit but still seems a little high indicating the presence of disorder. While disorder may give rise to interesting physical phenomena from basic physics point of view, it is desirable to control the amount of disorder and consequent residual resistivity from both basic and application points of view. The origin of disorder in the SrIrO$_3$ system could be intrinsic or extrinsic considering that we are dealing with thin films. Although we gave lots of efforts to optimize the growth conditions, experimental improvements are still desired. X-ray diffraction measurements showed that disorder in the films increases with increasing strain resulting in broadening of the Bragg peak. SIMS data revealed chemical inhomogeneity arising from elemental mixing at the film-substrate interface as a possible origin. Thickness dependent FWHM indicated the presence of structural defects above certain thickness. Also a possibility of an impurity phase cannot be completely ruled out because perovskite SrIrO$_3$ is in fact a metastable phase. Oxygen vacancies in oxide films can never be neglected as a source of atomic disorder. Identifying and eliminating all these possible origins of disorder would be a challenging task and require extensive efforts in detailed microscopic analysis by transmission electron microscopy, x-ray diffraction etc [36,37].

In closing, it is admitted that the present work is a commencement not completion of presumably time consuming efforts yet to come in clarifying the physics of SrIrO$_3$. In particular, the experimental methods employed for the characterization have been limited to macroscopic transport measurements, and none of the microscopic or spectroscopic tools such as, for example, optical conductivity measurement and angle resolved photoemission spectroscopy (ARPES), have been utilized [38,39]. Currently measurements along these directions are in progress.

**ACKNOWLEDGEMENT**

We would like to thank Prof. J. -S. Kim, Prof. H. J. Lee, S. Y. Kim, Y. W. Lee, S. W. Kim, Dr. N. Das for technical helps and discussions. AB would like to thank KBSI for the RSM measurements. YHJ thanks the Department of Physics, University of Virginia, where part of the work has been done, for the hospitality. The present work was supported by the National Research Foundation of Korea (2011-0009231, and 2012-R1A1B3000550) and SRC at POSTECH (2011-0030786).

**TABLE I**: In-plane and out-of-plane lattice parameters *a* (Å) and *c* (Å), respectively, and unit cell volume of 35 nm SrIrO$_3$ films grown on different substrates. The bulk lattice parameters are approximate ones converted from the orthorhombic ones. Full width half maxima (FWHM) of the (001)$_C$ rocking scans are also given.

| SrIrO$_3$ | *a* (Å) | *c* (Å) | Unit cell volume (Å$^3$) | FWHM (°) |
|---|---|---|---|---|
| SrIrO$_3$ (bulk) | ~3.96 | ~3.95 | ~61.942 | |
| on GdScO$_3$ | 3.96 | 3.99 | 62.570 | 0.041 |
| on DyScO$_3$ | 3.94 | 4.01 | 62.249 | 0.042 |
| on SrTiO$_3$ | 3.90 | 4.03 | 61.296 | 0.044 |
| on NdGaO$_3$ | 3.86 | 4.06 | 60.492 | 0.046 |

**Figure Captions:**

**FIG. 1.** (Color Online) Pseudocubic lattice parameters of $SrIrO_3$ and various substrates used in this study. $SrIrO_3$ and the substrates $GdScO_3$, $DyScO_3$, and $NdGaO_3$ are orthorhombic while $SrTiO_3$ is cubic. The orthorhombic substrates have the (110) orientation and the cubic one does (001).

**FIG. 2.** (Color Online) (a)-(d) X-ray $\theta - 2\theta$ scan results of $SrIrO_3$ films of thickness 35 nm grown on different substrates, $GdScO_3$ (110), $DyScO_3$ (110), $SrTiO_3$ (001) and $NdGaO_3$ (110). $SrIrO_3$ film and underlying substrate peaks are seen. Note that the film peak is cubic-indexed as $(001)_C$. (e) Rocking curves ($\omega$ scans) around the $(001)_C$ peak of $SrIrO_3$ films of various thicknesses on $GdScO_3$. Zero value of $\Delta\omega$ corresponds to the maximum peak position. As the thickness increases, full width at half maximum (FWHM) stays small up to 40 nm but becomes broad above that. FWHM for different thicknesses is indicated in the figure. (f) Cross sectional TEM image of $SrIrO_3$ grown on $GdScO_3$ substrate reveals well matched atomic alignment. (g) Reciprocal space mapping (RSM) of $SrIrO_3$ $(103)_C$ peak reveals in-plane compressive strain in the films. While in-plane $Q_x$ of the films is locked to that of the substrates, out-of-plane $Q_z$ of the films decreases accordingly (out-of-plane expansion). (h)-(k) Secondary ion mass spectrometry (SIMS) reveals the in-depth distribution of the various ions of 35 nm $SrIrO_3$ films. While the films on $GdScO_3$, $DyScO_3$, and $SrTiO_3$ substrates have almost uniform distribution of Sr, Ir, and O, the film on $NdGaO_3$ shows an inhomogeneous distribution of Ir. AFM images of the $SrIrO_3$ films are also shown in the inset. All the films have flat surfaces with roughness of ~1 nm.

**FIG. 3.** (Color Online) (a) Temperature dependence of the electrical resistivity of a 35 nm SrIrO$_3$ film grown on GdScO$_3$. (b) The resistivity is plotted against $T^{4/5}$ to reveal the unusual sublinear dependence from 10 K up to 300 K. Inset illustrates that the resistivity shows Fermi liquid behavior ($\propto T^2$) below 10 K down to 300 mK.

**FIG. 4.** (Color Online) Transport behaviors of 35 nm SrIrO$_3$ film on GdScO$_3$. (a) Mobility and carrier density, obtained from Hall measurements, are shown as a function of temperature. (b) Comparison of resistivity with and without a field of 9 Tesla indicates an appearance of magnetoresistance (MR) at low temperatures. (c) In-plane and out-of-plane transverse MRs, indicating a three dimensional character, are shown.

**FIG. 5.** (Color Online) (a) Temperature dependence of the sheet resistance of SrIrO$_3$ films on GdScO$_3$ having various thicknesses. The vertical axis is in logarithmic scale. The parallel line indicates the Mott-Ioffe-Regel (MIR) limit of quantum resistance, $h/e^2 \approx 26$ k$\Omega$, and the sheet resistance of the 3 nm film crosses the line. Low temperature upturn appears for the 4 nm film as indicated by an arrow. (b) Sheet conductance (inverse of sheet resistance) of the 4 nm film in the upturn region is plotted against ln $T$ from two dimensional weak localization model. (c) For the 3 nm insulating film, low temperature conductivity follows two dimensional variable range hopping well up to approximately 15 K.

**FIG. 6.** (Color Online) (a)-(c) The sheet resistance ($R_S$) of metallic SrIrO$_3$ films on GdScO$_3$ having different thicknesses is plotted against $T^{4/5}$. They all show the unusual sublinear temperature dependence. (d) Magnetoresistance (MR), *[R(B) - R(0)]/R(0)*, for the film of thickness 4 nm at various temperatures. At 5 K there appears a strong negative MR in contrast to positive MR at 20 K.

**FIG. 7.** (Color Online) Electrical resistivity of thickness $t = 35$ nm SrIrO$_3$ films grown on GdScO$_3$, DyScO$_3$, SrTiO$_3$, and NdGaO$_3$ are shown as a function of temperature. Different substrates amount to different degree of compressive stress and consequent strain in the films. Metal insulator transition is seen to occur as the compressive strain is increased (arrow in the figure). The films on GdScO$_3$ and NdGaO$_3$ are metallic and insulating, respectively. Films with intermediate strains, grown on DyScO$_3$ and SrTiO$_3$, are metallic but low temperature upturns in resistivity start to appear as temperature is decreased. Arrows in the figure indicate minima which occur at 20 K (DyScO$_3$) and 50 K (SrTiO$_3$).

**FIG. 8.** (Color Online) (a) Resistivity of SrIrO$_3$ deposited on GdScO$_3$ is proportional to $T^{4/5}$. Inset illustrates the Fermi liquid behavior ($\rho \propto T^2$) below 10 K. (b) Resistivity of SrIrO$_3$ on DyScO$_3$ is linear in temperature. Inset shows that the conductivity below 20 K is proportional to $\ln T$. (c) $T^{3/2}$ power law is observed in resistivity of SrIrO$_3$ grown in SrTiO$_3$. Inset again shows that the conductivity below 50 K is proportional to $\ln T$. (d) For insulating films on NdGaO$_3$, three dimensional variable range hopping describes conductivity adequately below 35 K. The thickness of all the four films is 35 nm.

**FIG. 9.** (Color Online) (a)-(d) Magnetoresistance (MR) of SrIrO$_3$ films of $t = 35$ nm on GdScO$_3$, DyScO$_3$, SrTiO$_3$ and NdGaO$_3$ are shown. The magnetic field was in the out-of-plane direction. MR of all the films is positive and proportional to $B^2$.

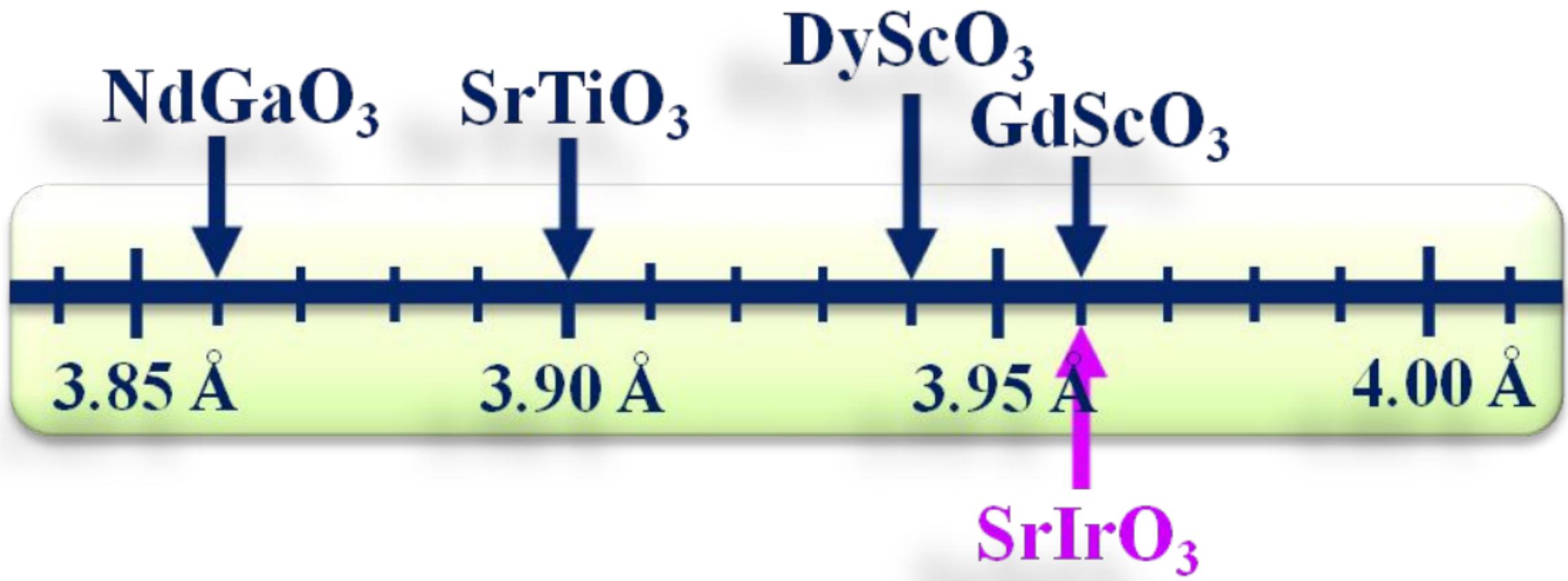

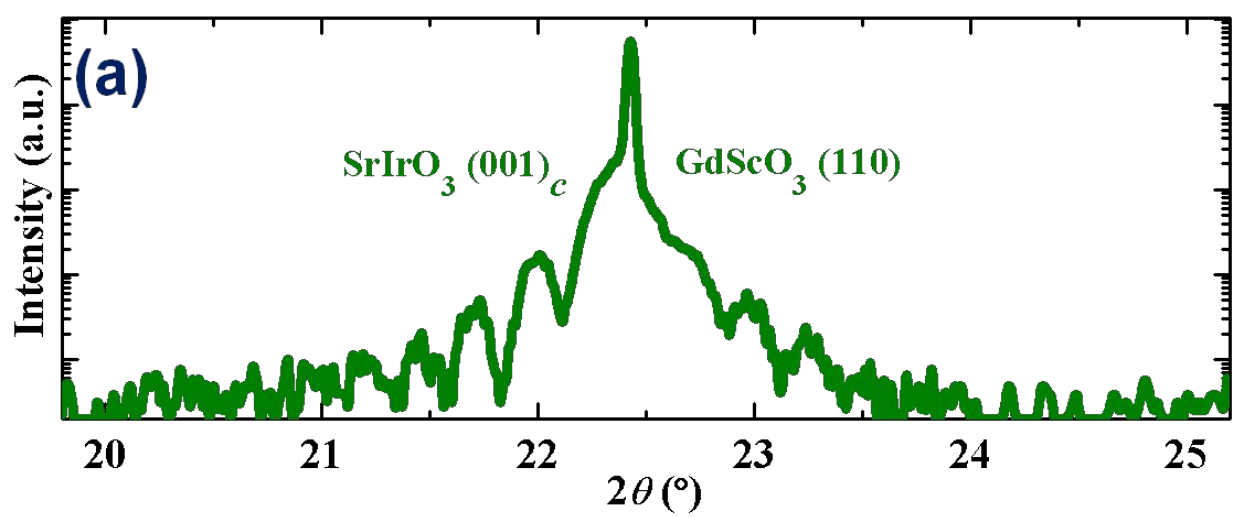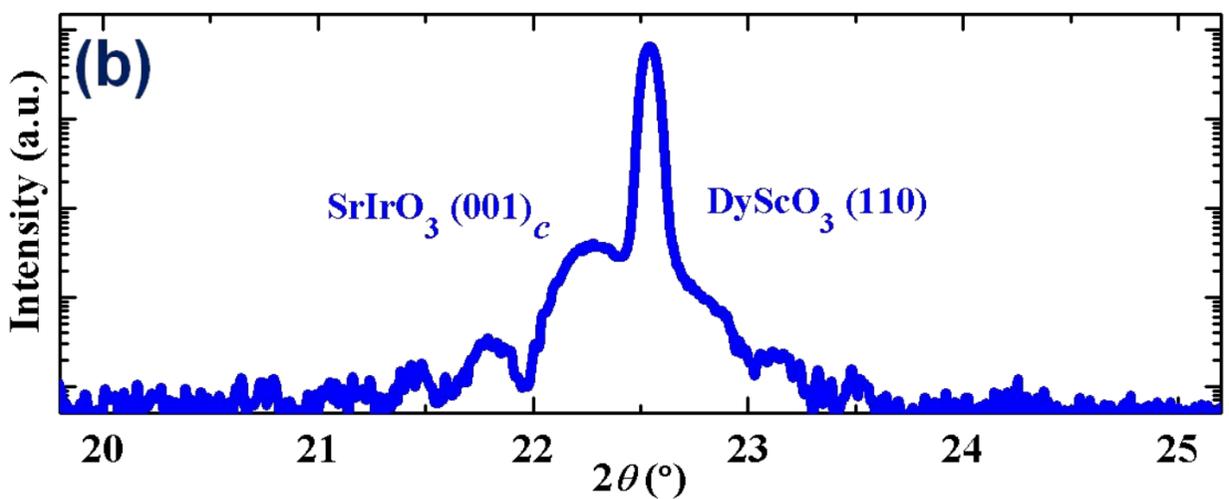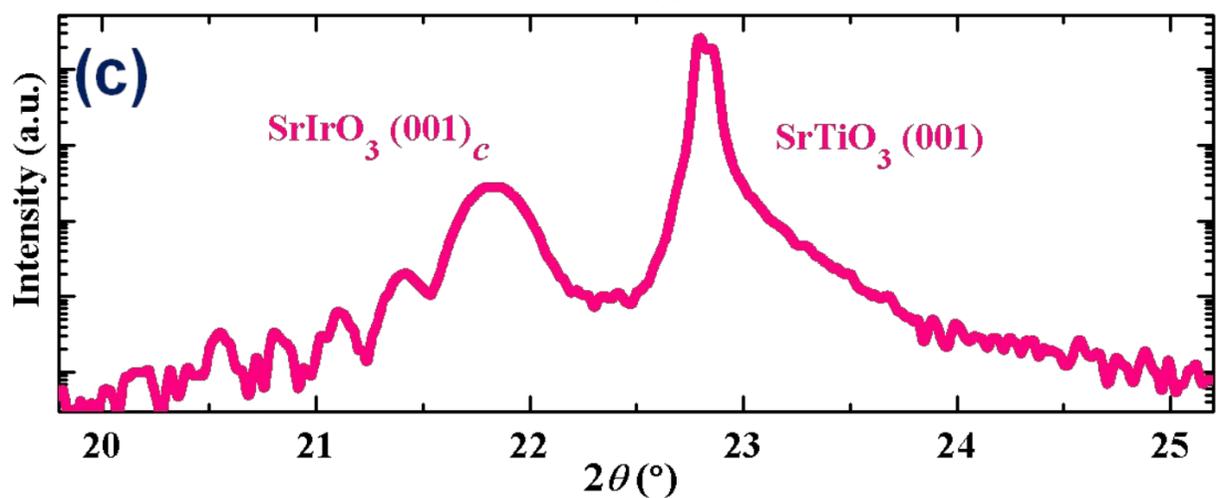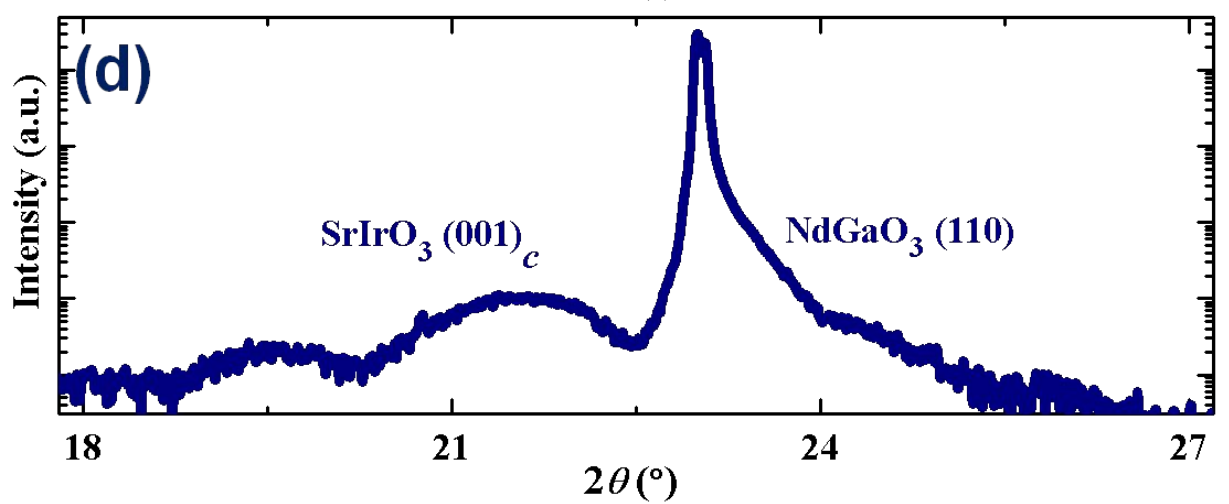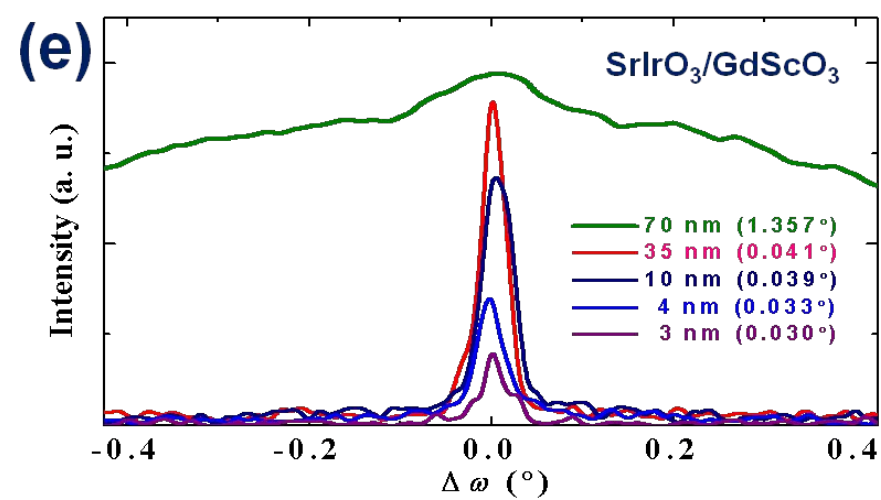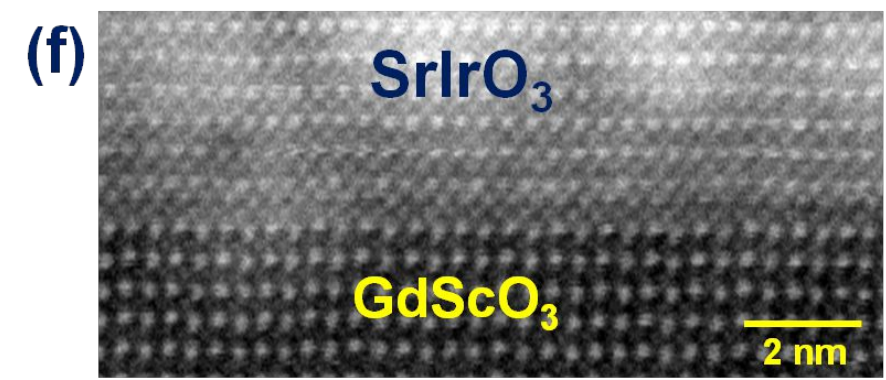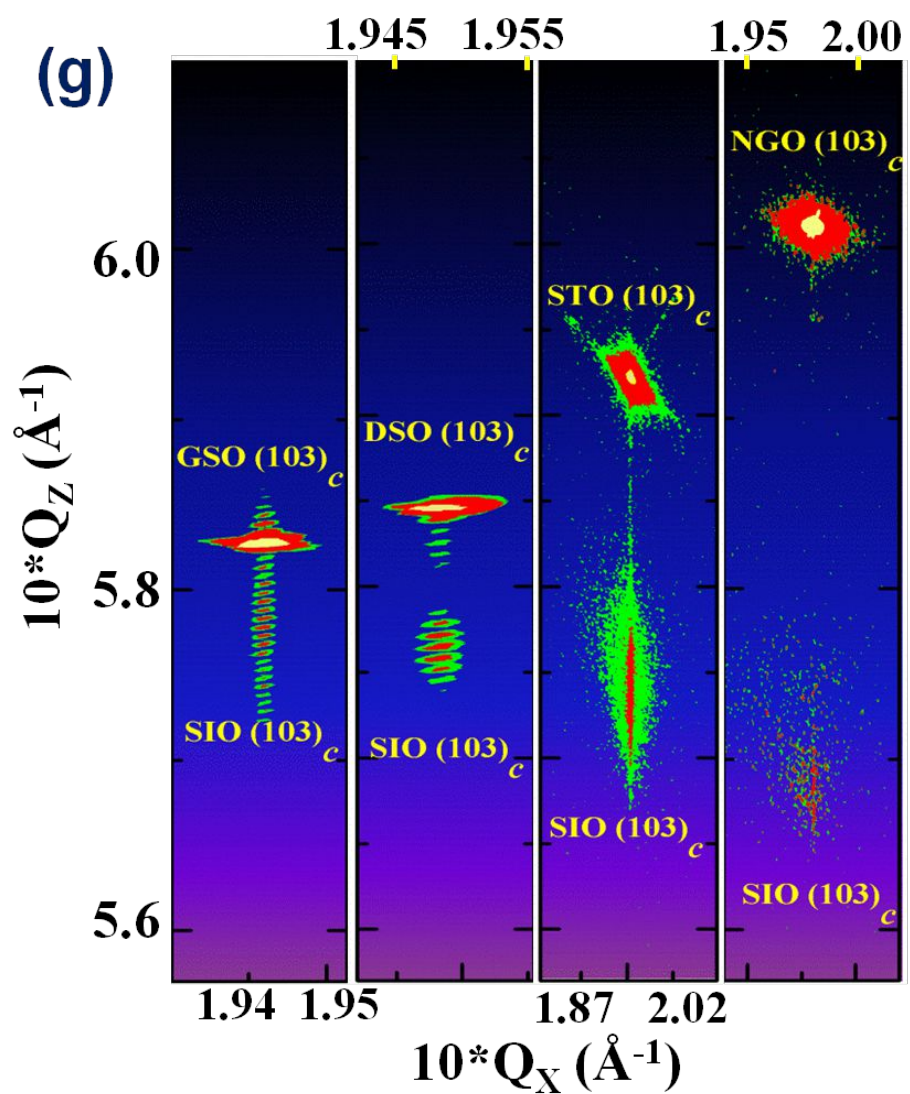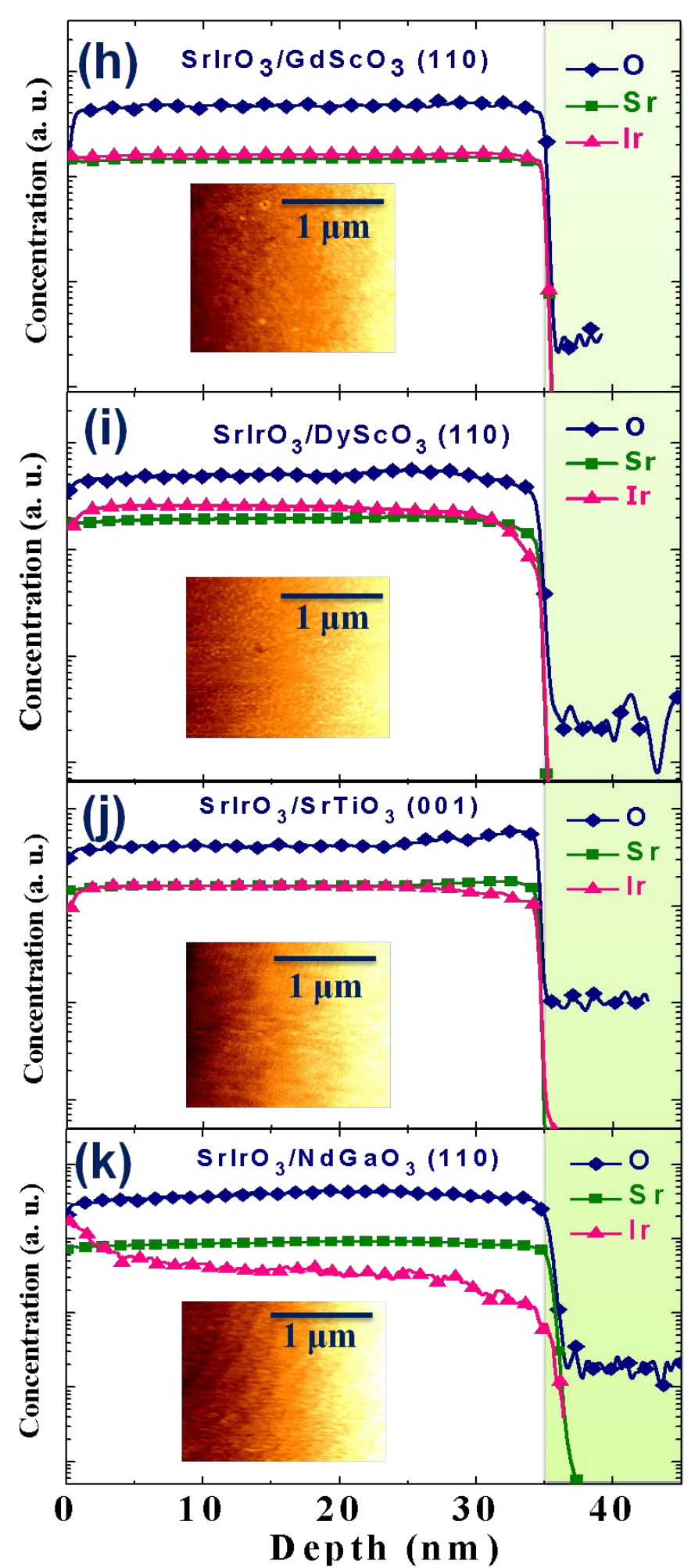

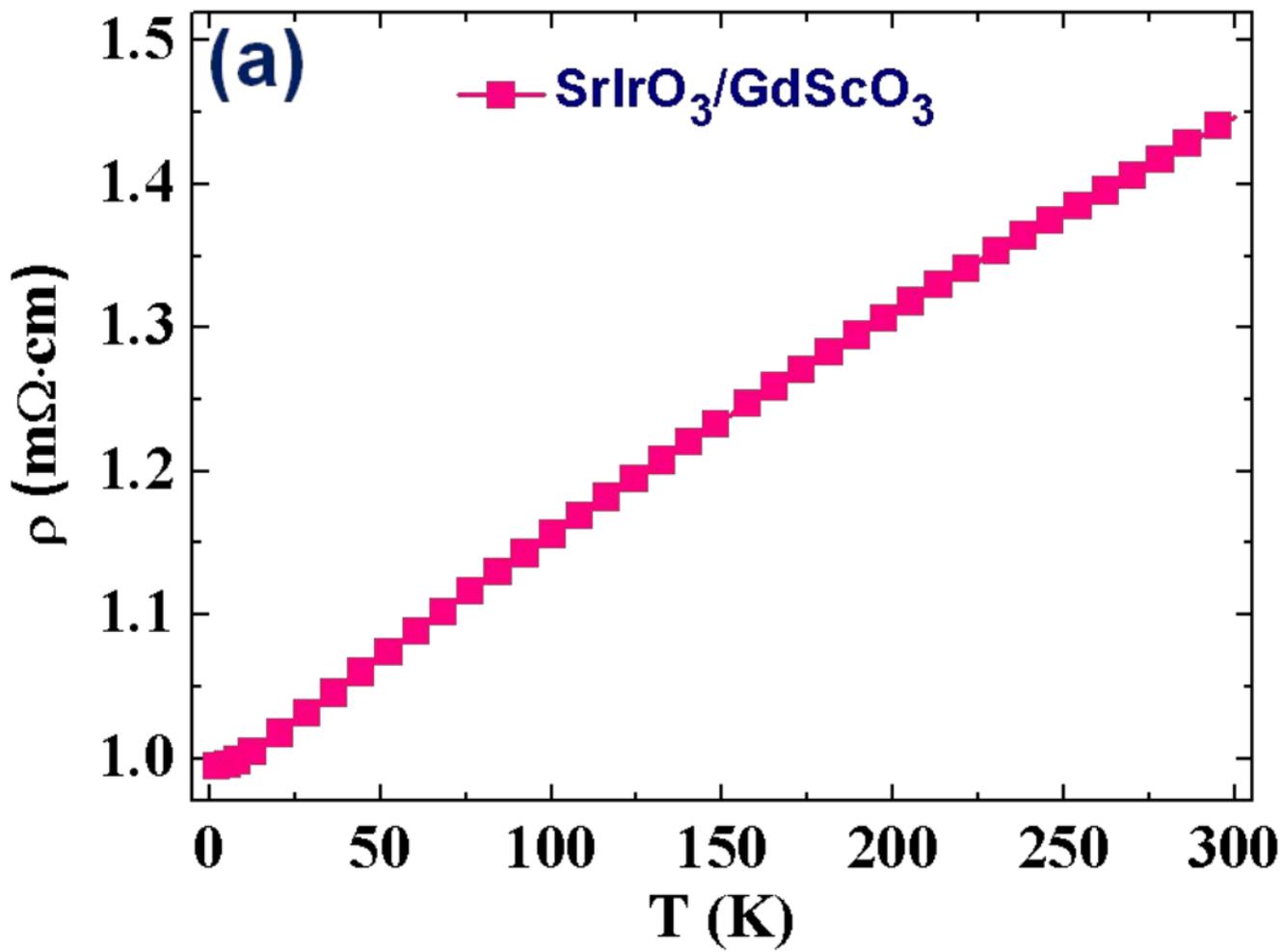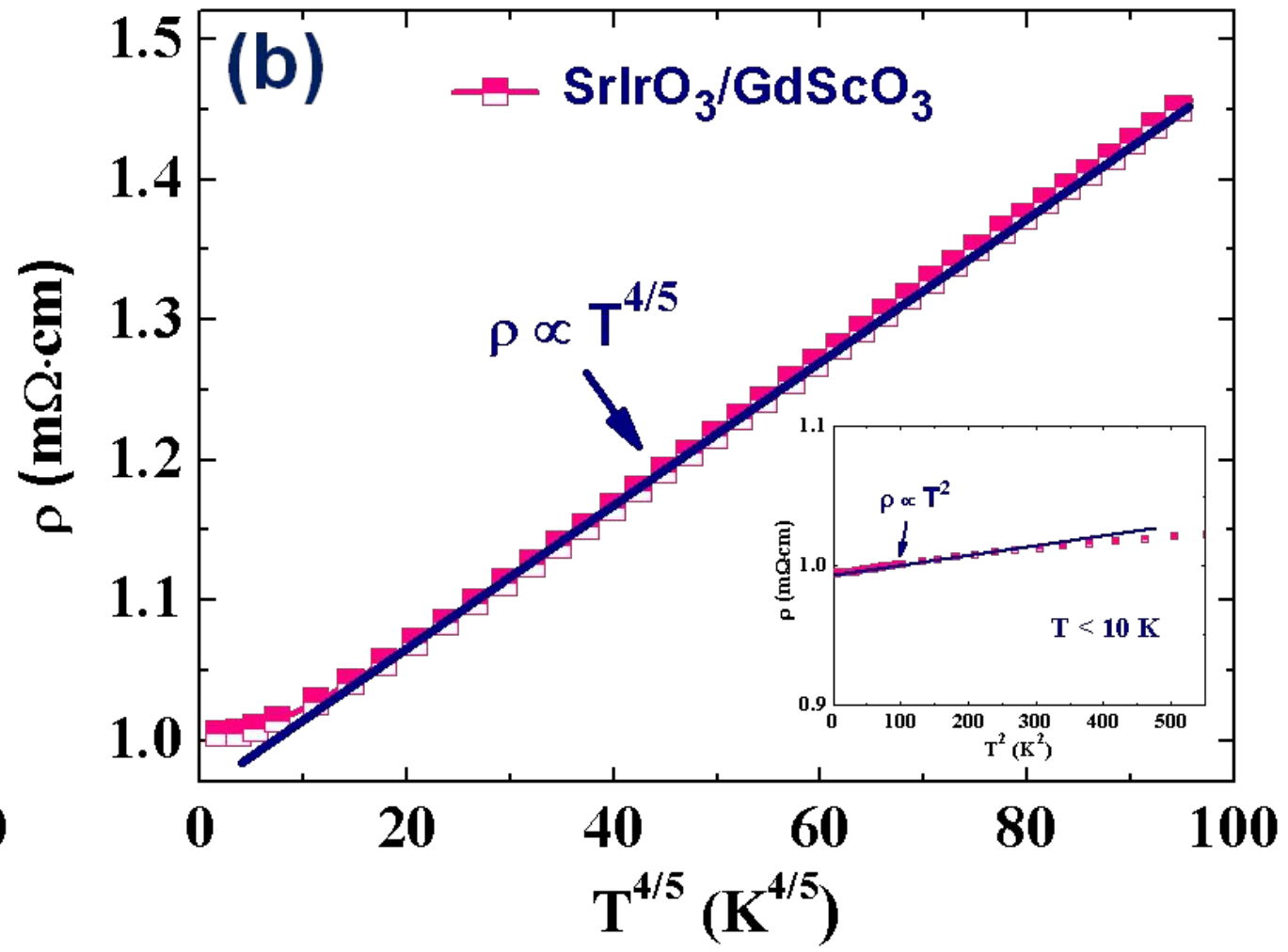

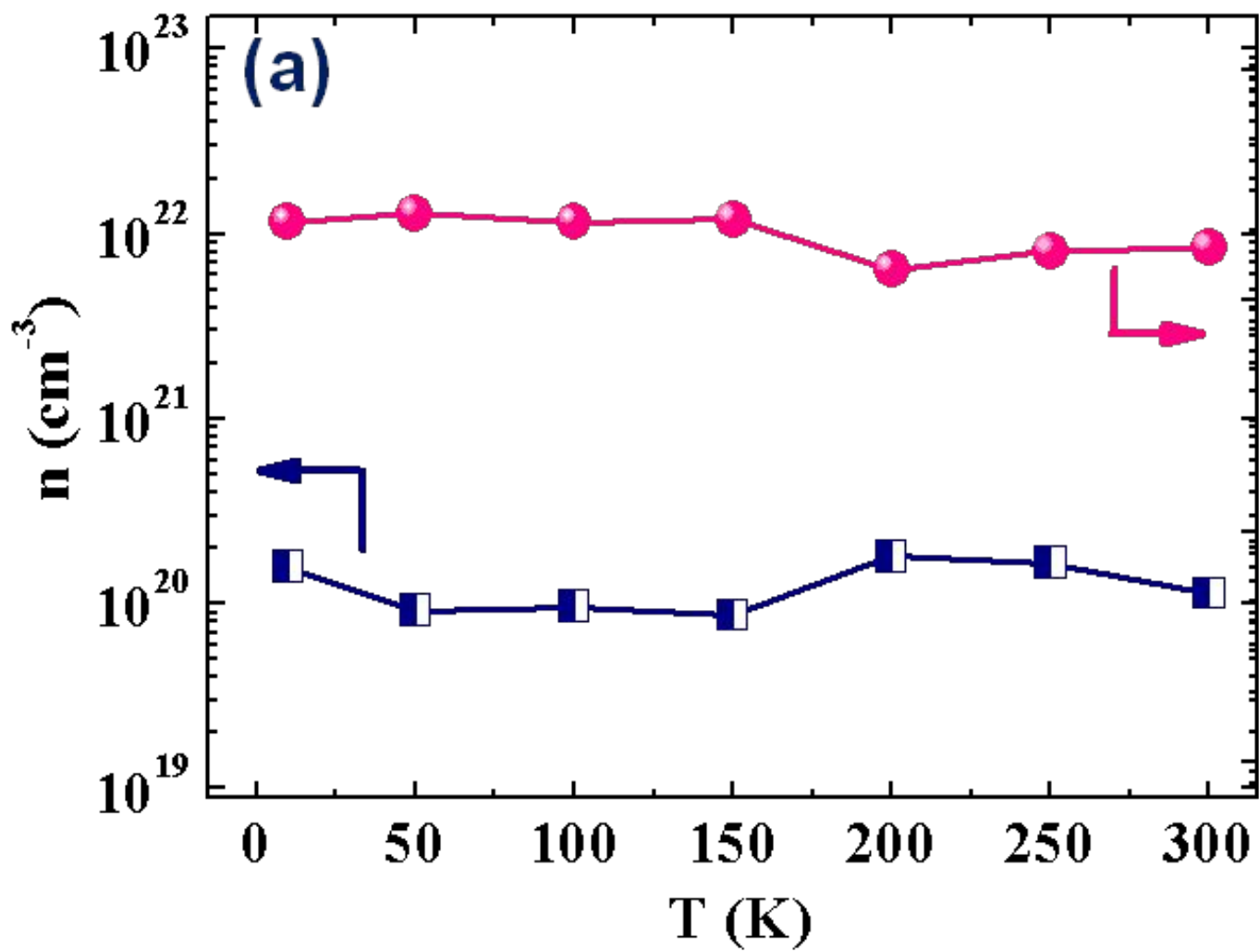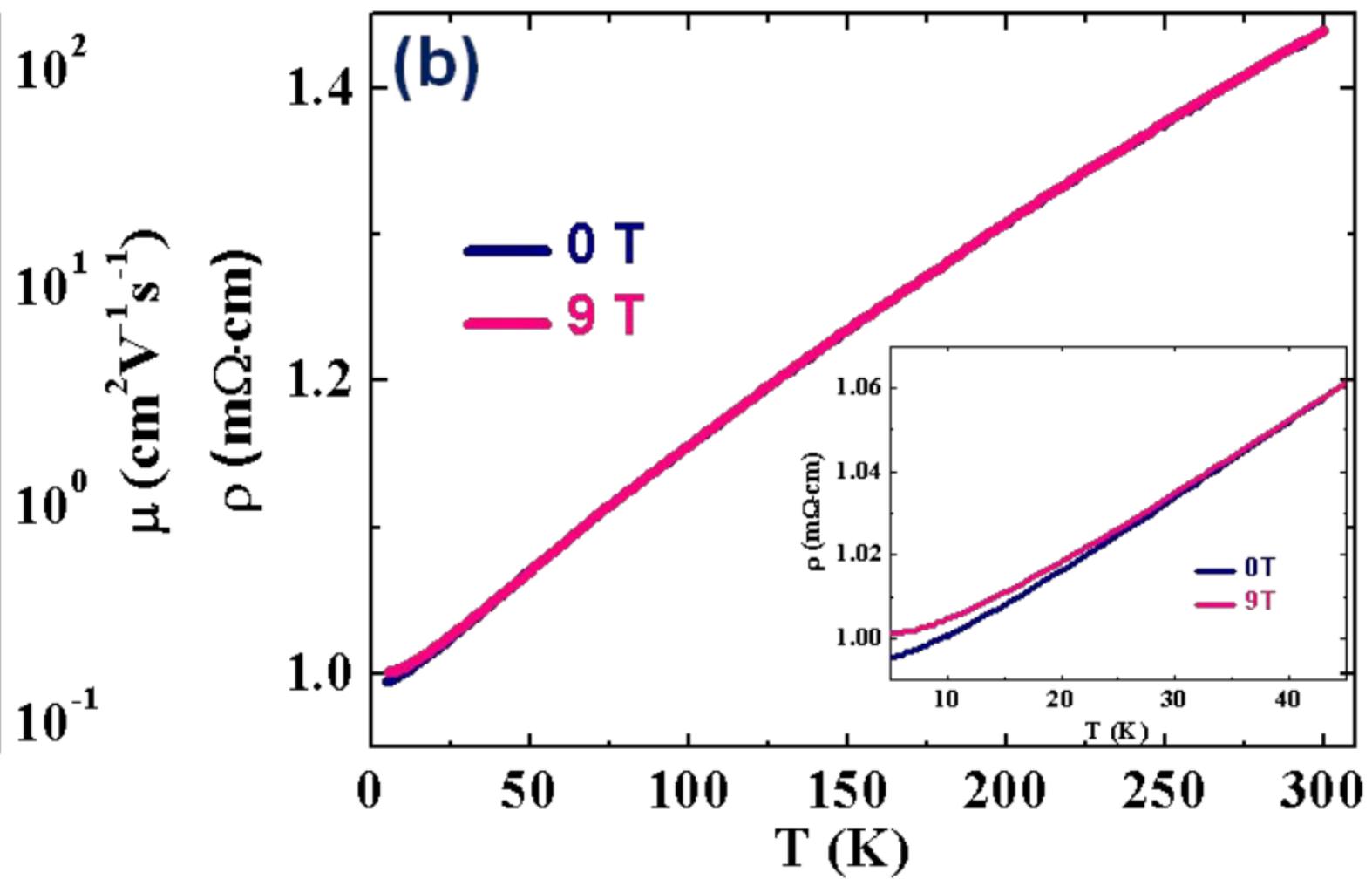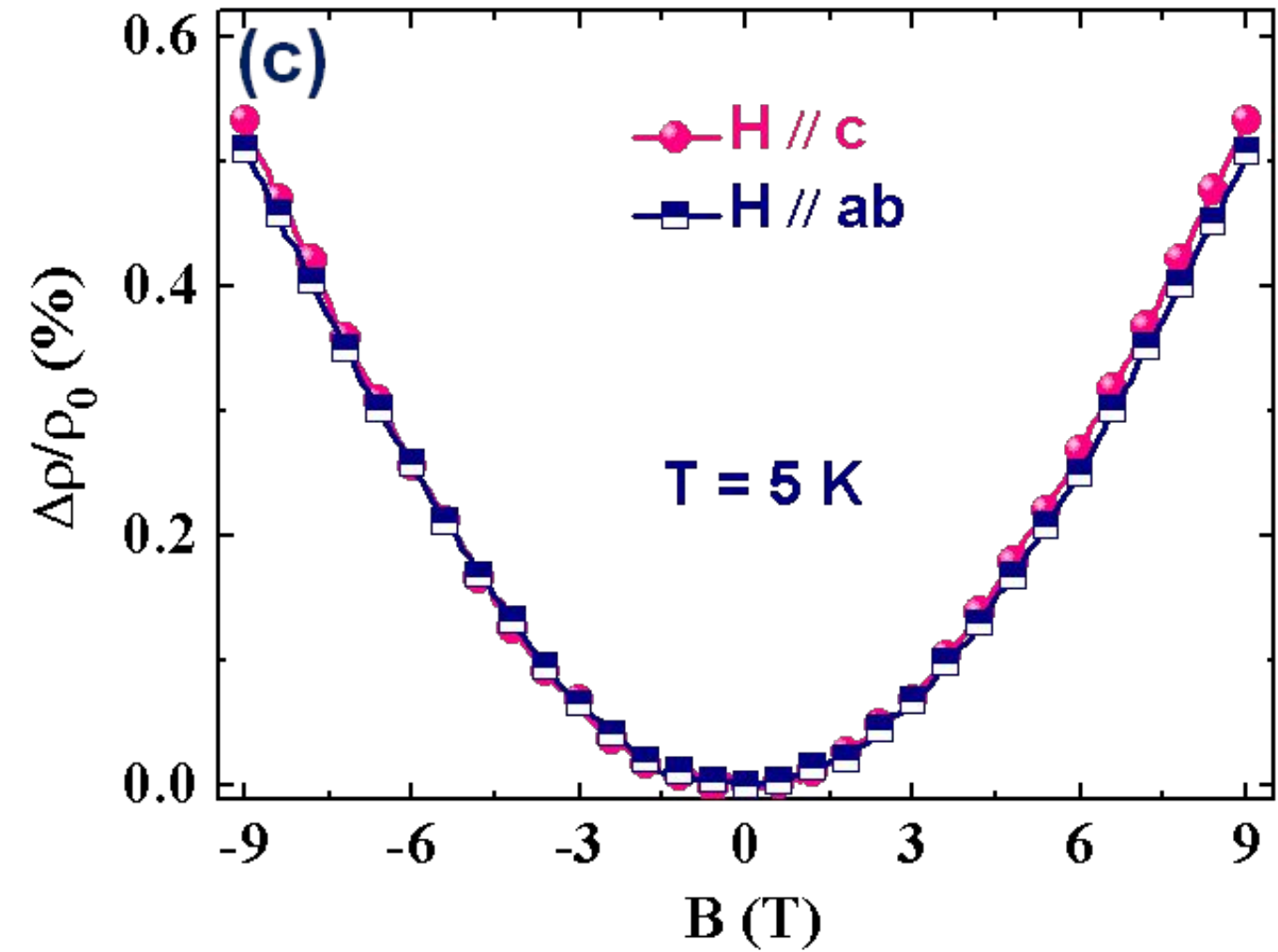

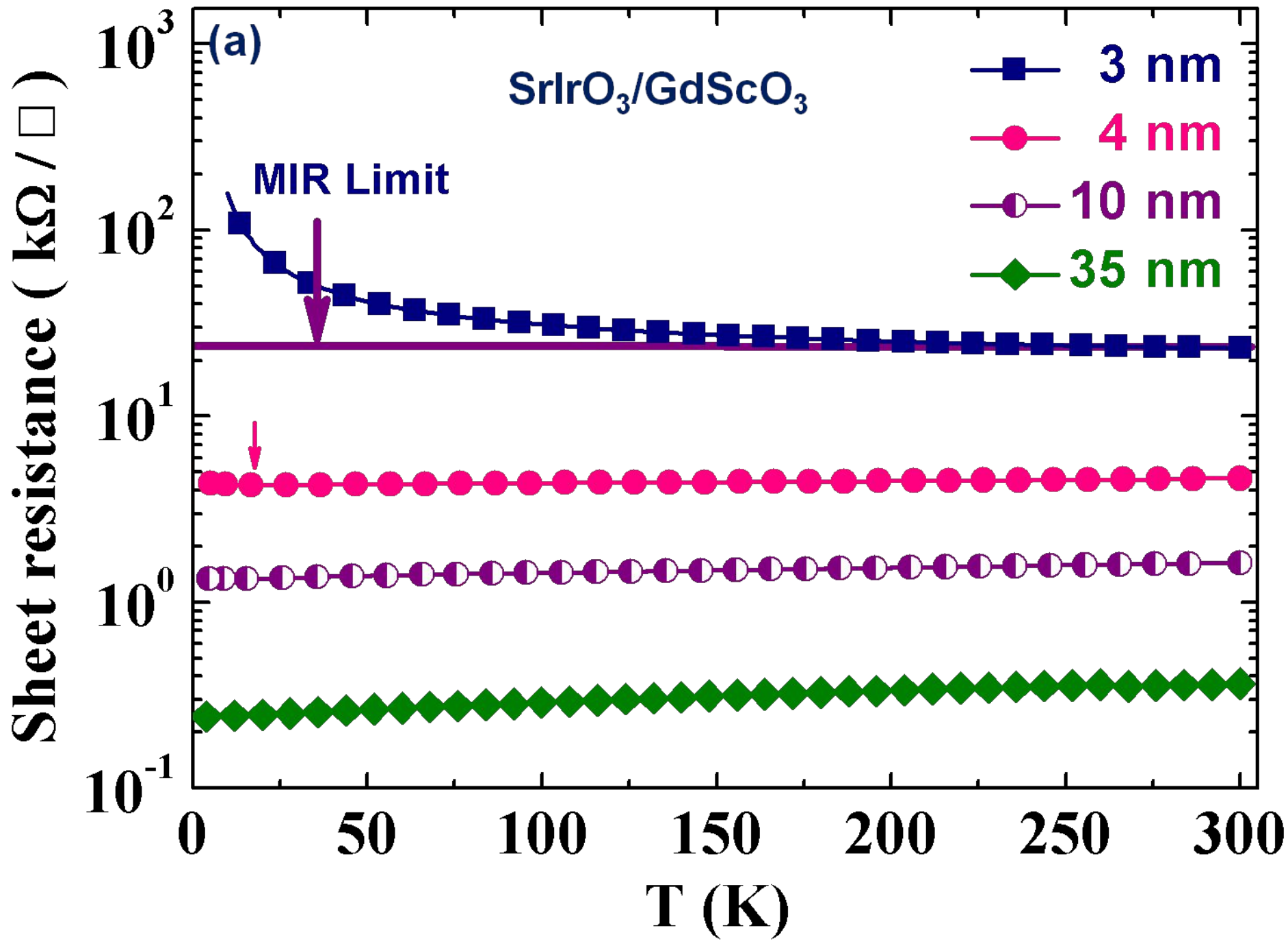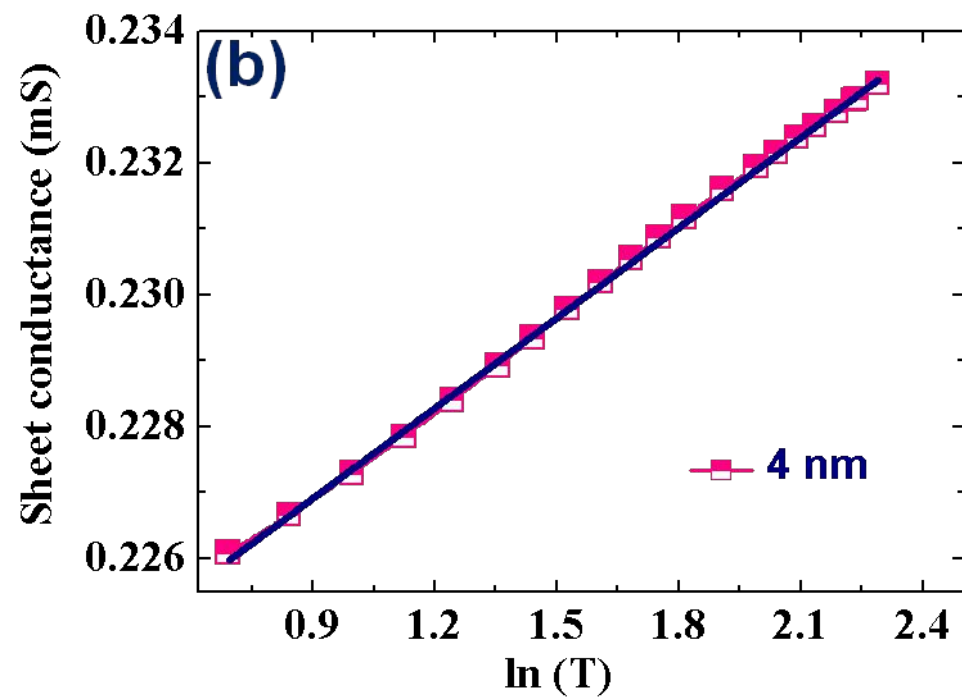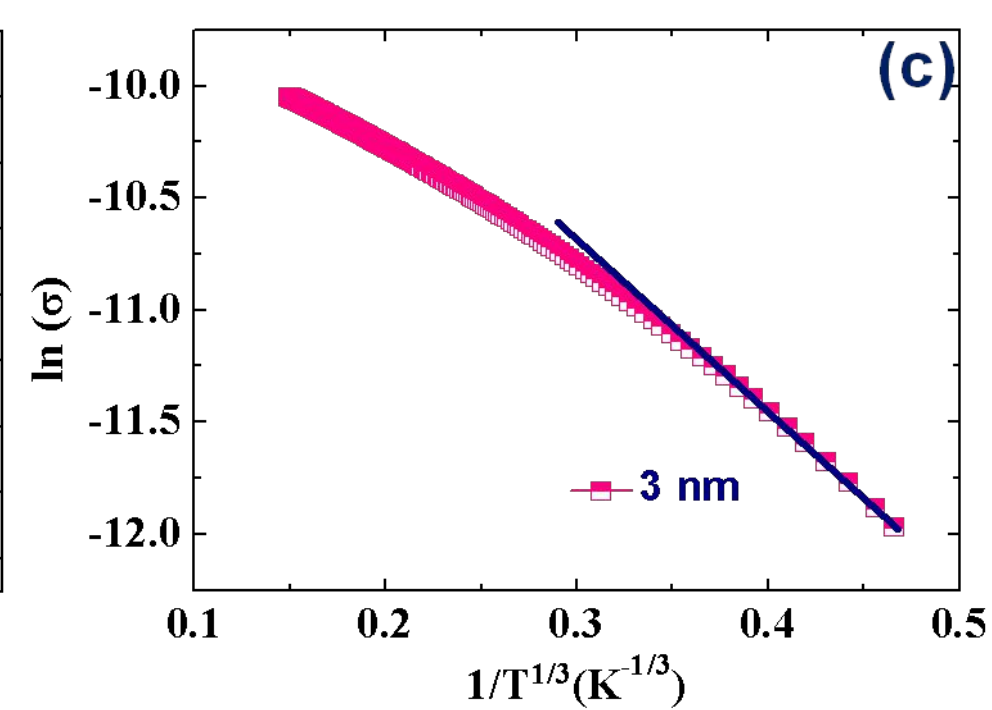

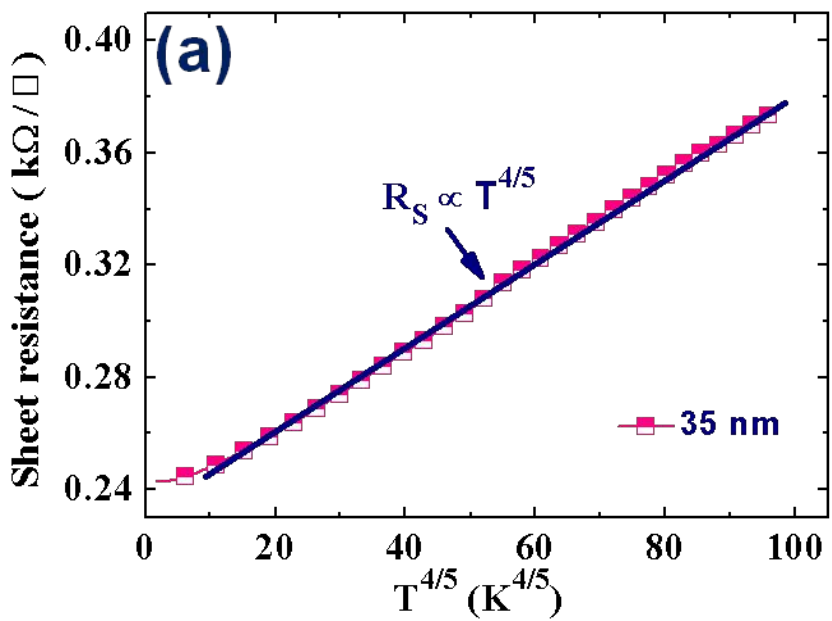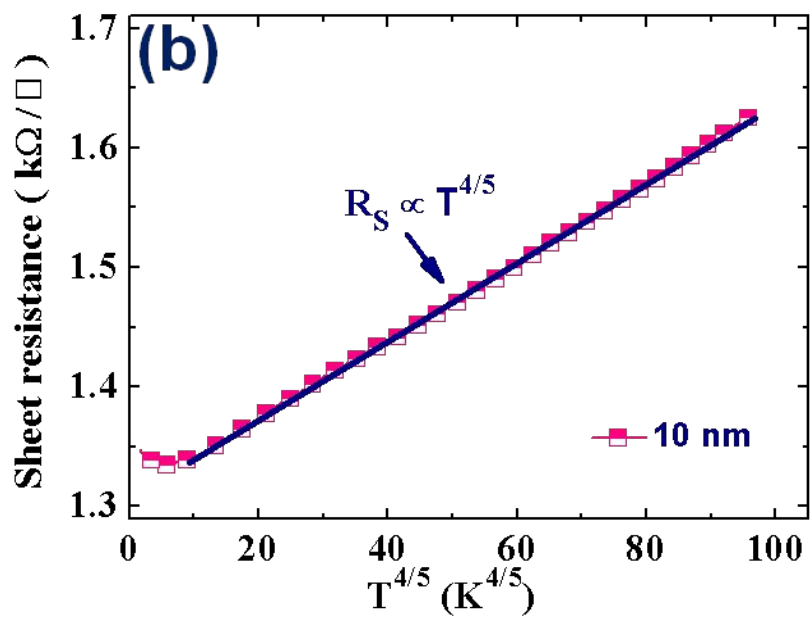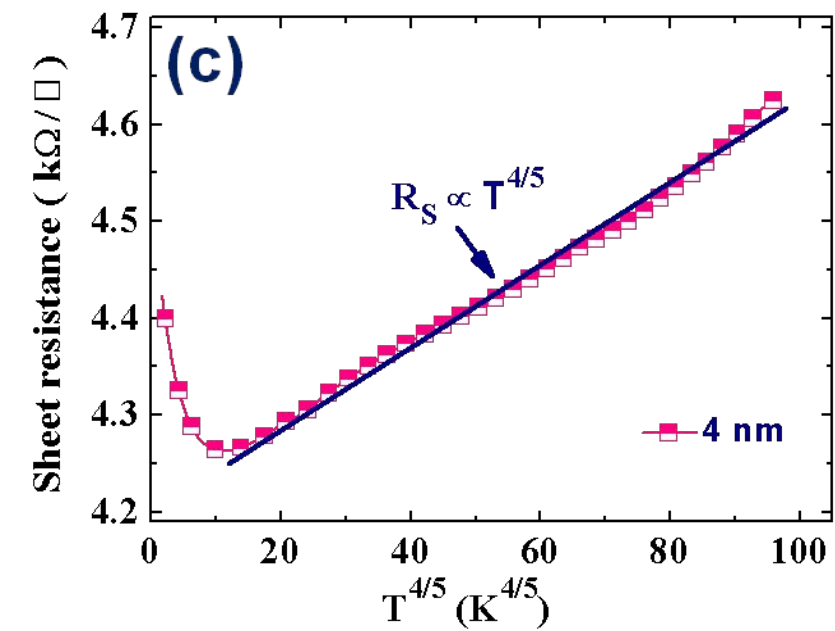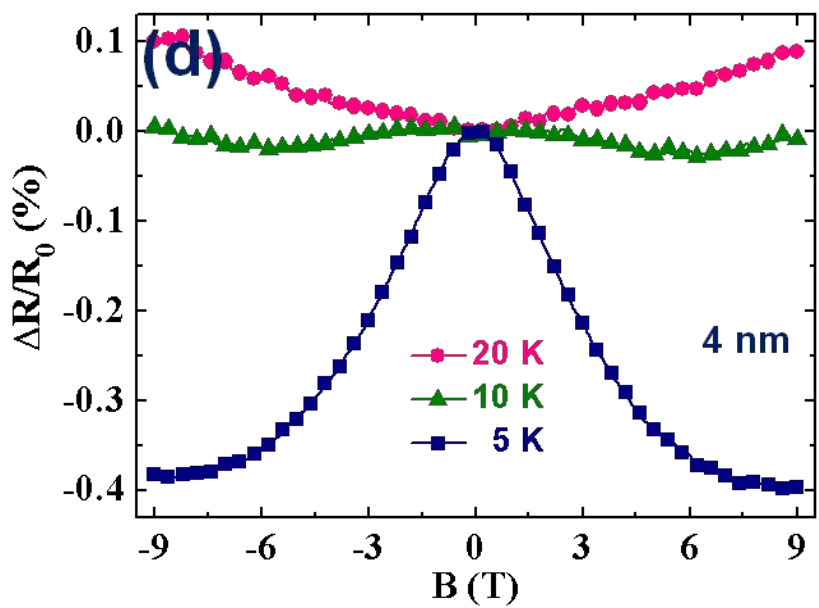

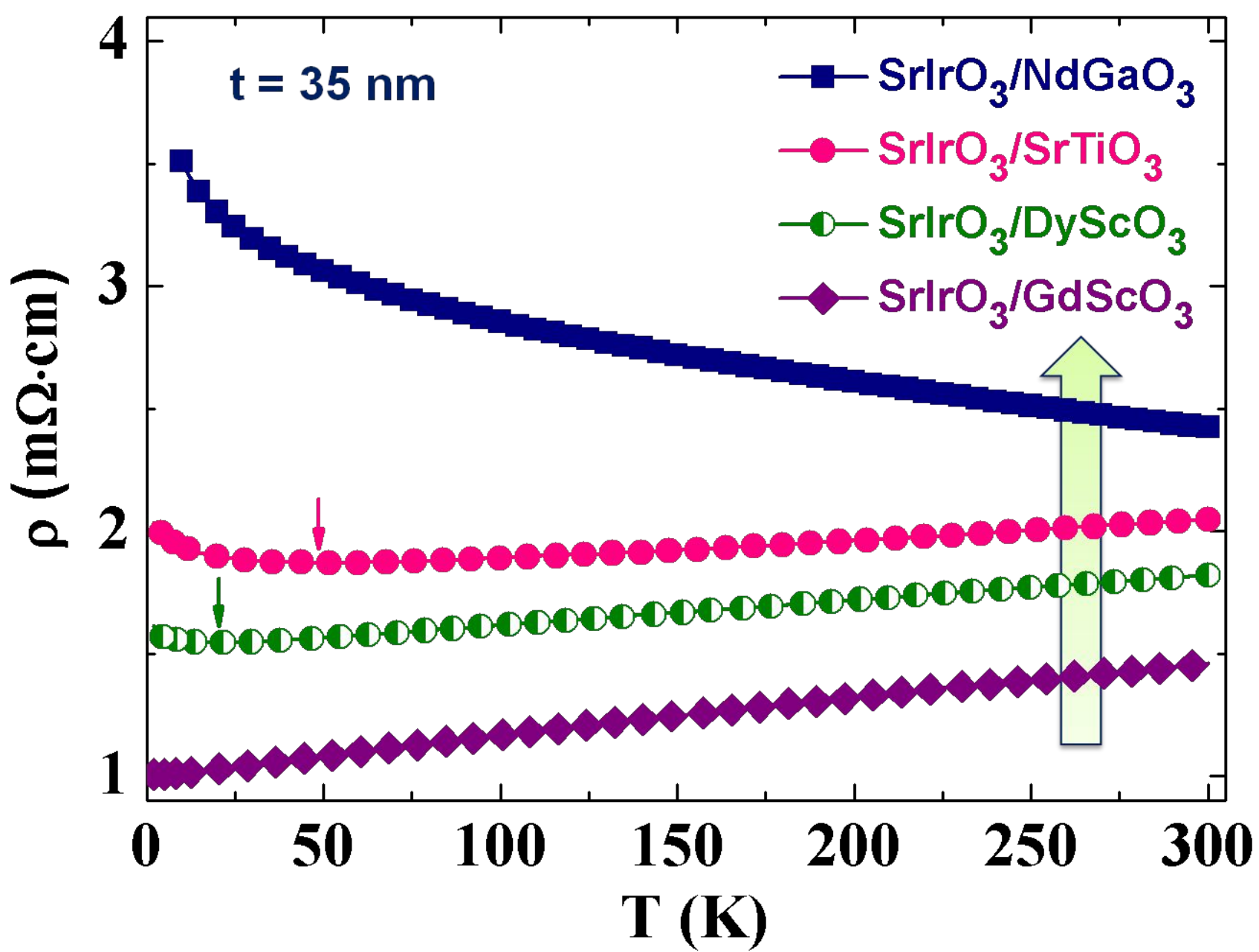

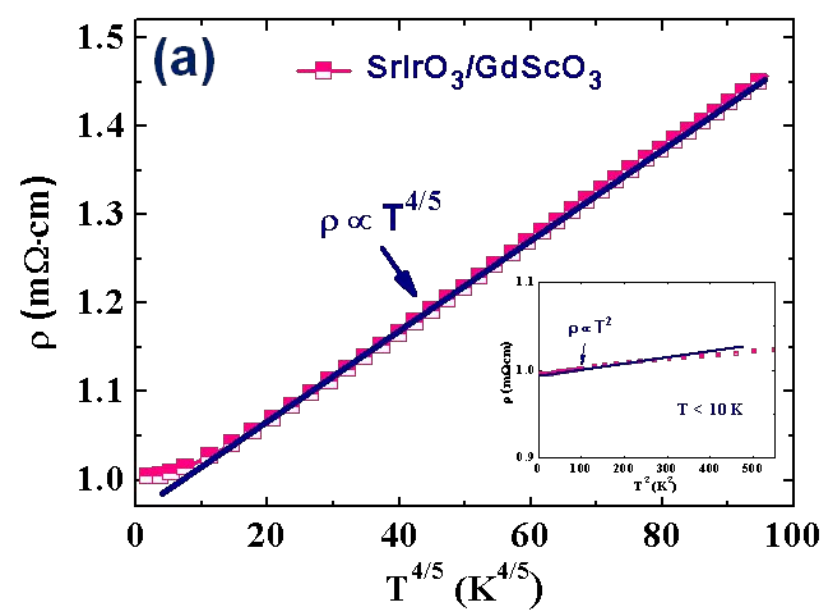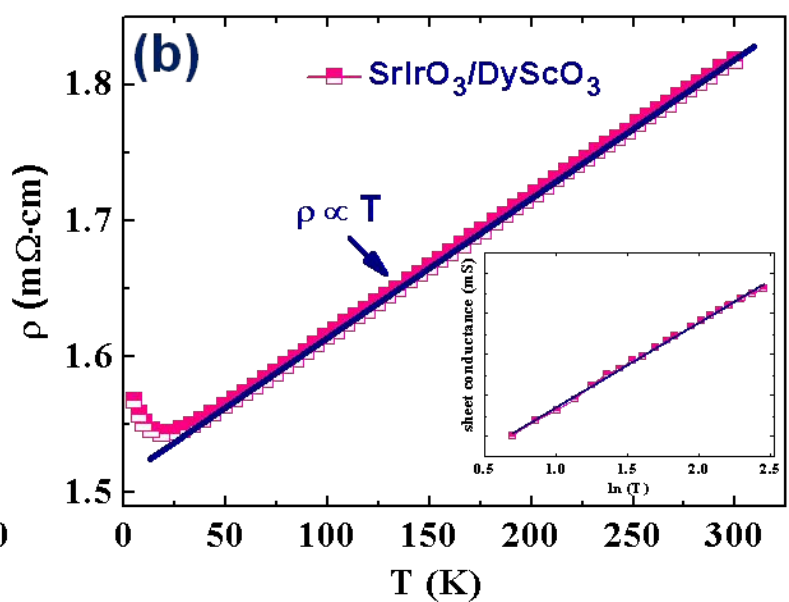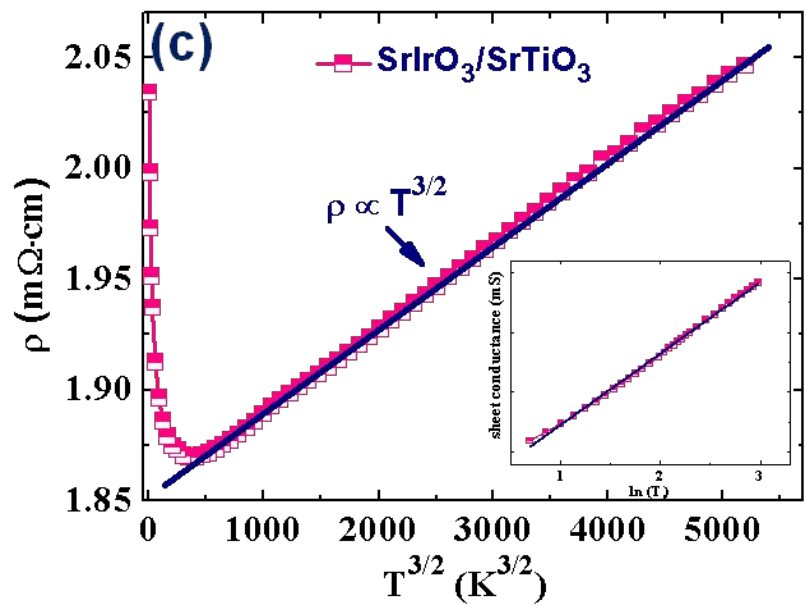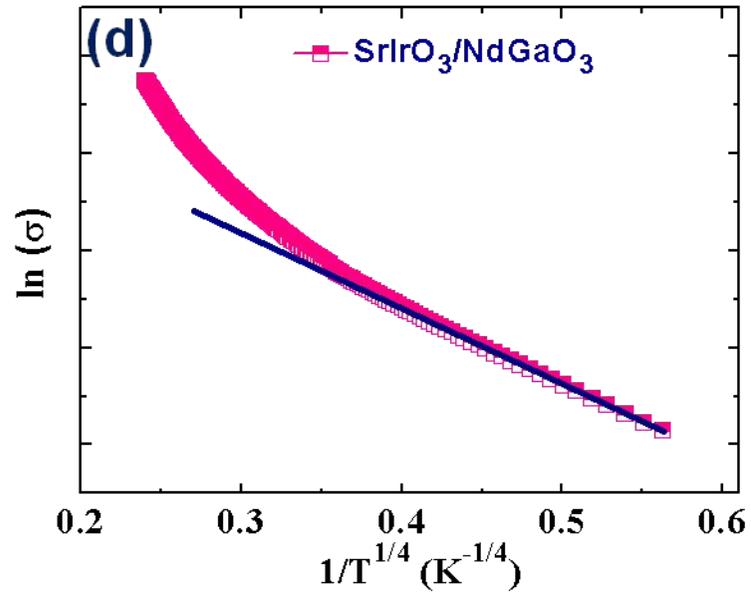

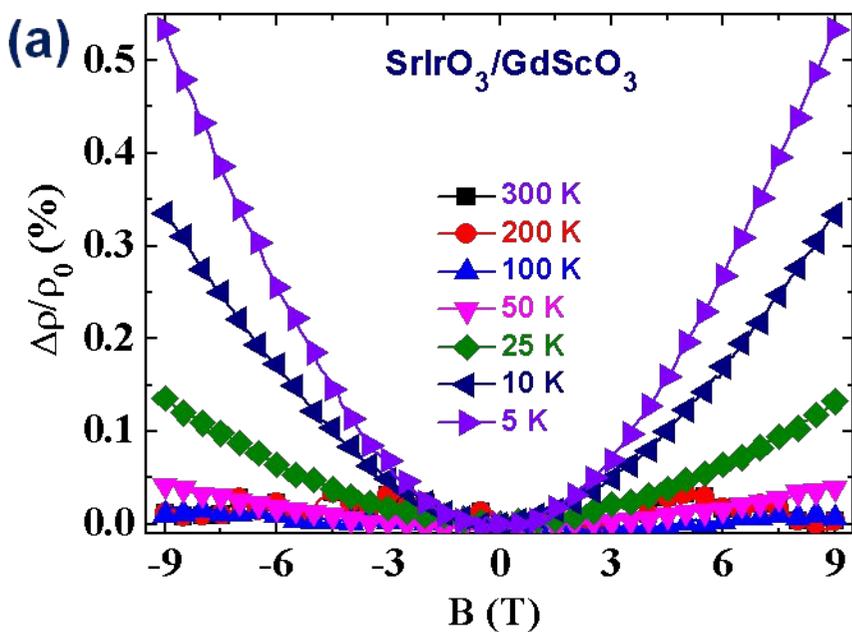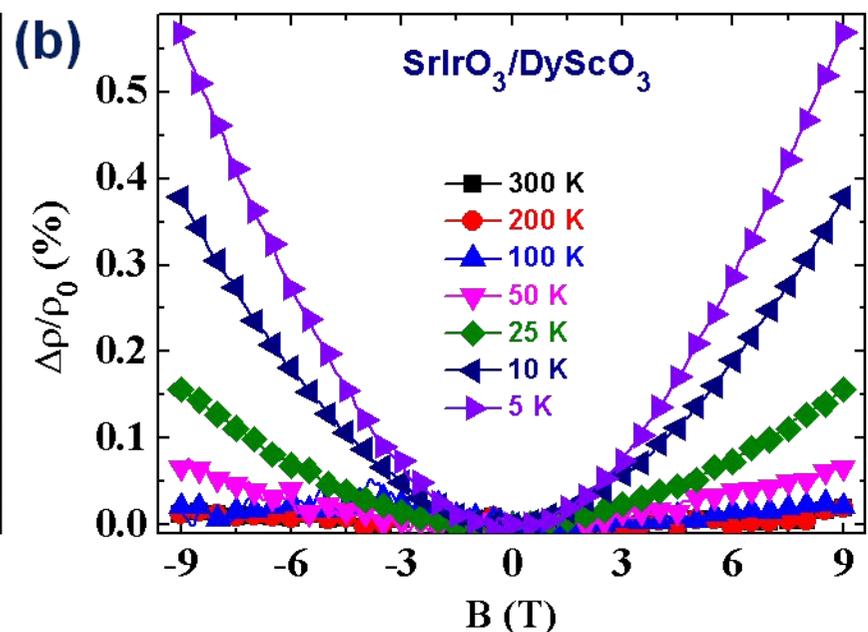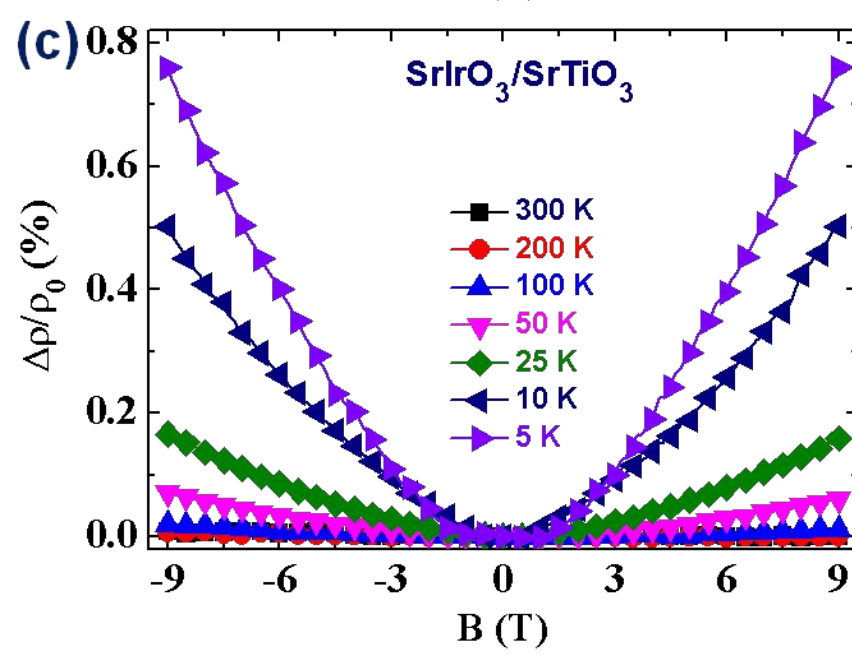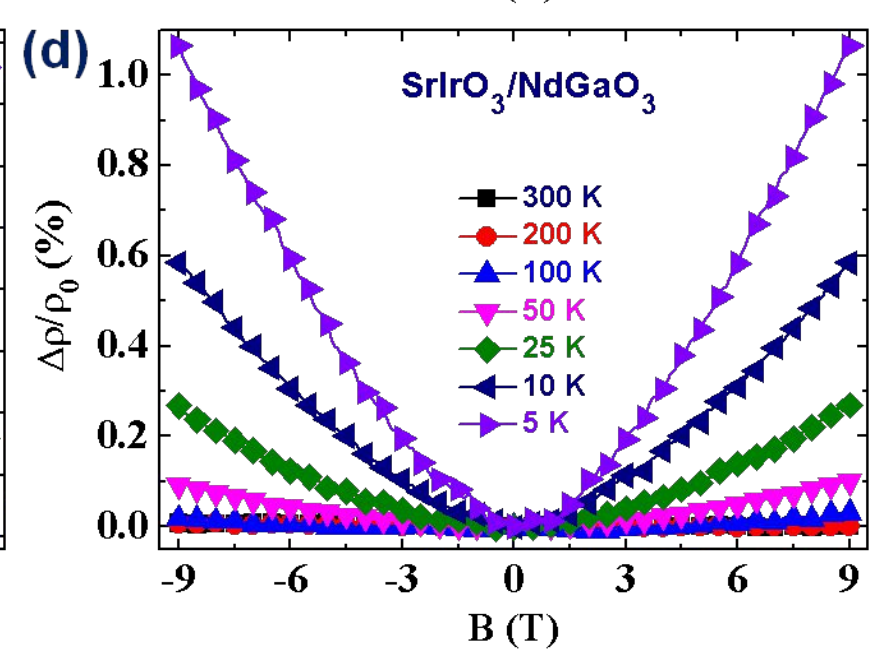